\documentclass[conference]{IEEEtran}

\usepackage{amsmath,amssymb,amsfonts}
\usepackage{algorithmic}
\usepackage{graphicx}
\usepackage{textcomp}
\usepackage{xcolor}
\usepackage{todonotes}
\usepackage{relsize}
\usepackage[final]{changes}

\usepackage[style=ieee,
bibstyle=ieee,
backend=biber,
minnames=1,
maxcitenames=2,
maxbibnames=6,
doi=true,
isbn=false,
url=false,
natbib=true,
date=year,
]{biblatex}

\usepackage{xpatch}

\xpatchbibdriver{misc}
  {\printfield{type}}
  {%
    \printfield{type}%
    \setunit{\addspace}%
    \printfield{shorttitle}%
  }
  {}{}

\AtEveryBibitem{%
  \clearfield{note}%
  \clearfield{langid}%
  \clearfield{language}%
  \clearlist{language}%
}

\usepackage{xpatch}
\usepackage{xstring}

\DeclareSourcemap{
	\maps[datatype=bibtex]{
		\map{
			\step[fieldsource=langid, match=\regexp{\A(n)?((swiss)?german|austrian)\Z}, final]
			\step[fieldset=language, fieldvalue={(in German)}]
		}
		\map{
			\step[fieldsource=langid, match=pinyin, final]
			\step[fieldset=language, fieldvalue={(in Chinese)}]
		}
		\map{
			\step[fieldsource=langid, match=japanese, final]
			\step[fieldset=language, fieldvalue={(in Japanese)}]
		}
		\map{
			\step[fieldsource=langid, match=ko, final]
			\step[fieldset=language, fieldvalue={(in Korean)}]
		}
	}
}

\xpatchbibdriver{inproceedings}					 	
	{\usebibmacro{publisher+location+date}}			
	{\IfSubStr{\strfield{booktitle}}{\strfield{year}}{\usebibmacro{publisher+location}}{\usebibmacro{publisher+location+date}}} 
	{} 
	{\typeout{There was an error patching biblatex-ieee (specifically, ieee.bbx's @inproceedings driver)}} 

\newbibmacro*{publisher+location}{%
	\printlist{location}%
	\iflistundef{publisher}
	{\setunit*{\addcomma\space}}
	{\setunit*{\addcolon\space}}%
	\printlist{publisher}%
	\newunit}

\xpatchbibdriver{online}
{\printtext[parens]{\usebibmacro{date}}}
	{\iffieldundef{year}
		{}
		{\printtext[parens]{\usebibmacro{date}}}}
	{}
	{\typeout{There was an error patching biblatex-ieee (specifically, ieee.bbx's @online driver)}}

\DeclareSourcemap{
	\maps[datatype=biber]{
		\map{
			\step[fieldsource=note, final]
			\step[fieldset=addendum, origfieldval, final]
			\step[fieldset=note, null]
		}
	}
}

\DeclareBibliographyDriver{standard}{%
	\usebibmacro{bibindex}%
	\usebibmacro{begentry}%
	\usebibmacro{maintitle+title}%
	\setunit{\addcomma\addspace}%
	\usebibmacro{publisher+type+number}%
	\setunit{\labelnamepunct}\newblock
	\IfSubStr{\strfield{number}}{\strfield{year}}{}{\usebibmacro{date}} 
	\newunit\newblock
	\usebibmacro{finentry}%
}

\newbibmacro*{publisher+type+number}{%
	\printtext{%
		\printlist{publisher}
		\iflistundef{publisher}
		{\space}{}%
		\iffieldundef{type}
		{Standard}
		{\printfield{type}}
		\printfield{number}
	}%
}

\xpatchbibdriver{report}					 	
	{\usebibmacro{date}}			
	{\IfSubStr{\strfield{number}}{\strfield{year}}{}{\usebibmacro{date}}} 
	{} 
	{\typeout{There was an error patching biblatex-ieee (specifically, ieee.bbx's @report driver)}} 

\DeclareBibliographyDriver{software}{%
	\usebibmacro{bibindex}%
	\usebibmacro{begentry}%
	\usebibmacro{maintitle+title}%
	\setunit{\addcomma\addspace}%
	\usebibmacro{version+year}%
	\setunit{\labelnamepunct}\newblock
	\usebibmacro{author}%
	\setunit{\addcomma\addspace}%
	\newunit\newblock
	\usebibmacro{url+urldate}%
	\newunit\newblock
	\usebibmacro{finentry}%
}

\newbibmacro*{version+year}{%
	\printtext{%
		\iffieldundef{version}{\iffieldundef{year}{}{\printfield{year}}}{\printfield{version}}
		\printfield{number}
		\printlist{publisher}
		\iflistundef{publisher}{\space}{}%
	}%
}

\DeclareSourcemap{
	\maps[datatype=bibtex]{
		\map[overwrite, foreach={booktitle,journaltitle,eventtitle,series,publisher,institution,type}]{
			\step[fieldsource=\regexp{$MAPLOOP}, match={XXX}, replace={Acad. Serbe Sci. Arts Glas Cl. Sci. Tech.}]
			\step[fieldsource=\regexp{$MAPLOOP}, match={XXX}, replace={Acta Acust.}]
			\step[fieldsource=\regexp{$MAPLOOP}, match={XXX}, replace={Acta Astron. (Poland)}]
			\step[fieldsource=\regexp{$MAPLOOP}, match={XXX}, replace={Acta Astron. Sin. (China)}]
			\step[fieldsource=\regexp{$MAPLOOP}, match={XXX}, replace={Acta Astronaut. (U.K.)}]
			\step[fieldsource=\regexp{$MAPLOOP}, match={XXX}, replace={Acta Astrophys. Sin. (China)}]
			\step[fieldsource=\regexp{$MAPLOOP}, match={XXX}, replace={Acta Autom. Sin.}]
			\step[fieldsource=\regexp{$MAPLOOP}, match={XXX}, replace={Acta Cienc. Indica Math.}]
			\step[fieldsource=\regexp{$MAPLOOP}, match={XXX}, replace={Acta Cienc. Indica Phys.}]
			\step[fieldsource=\regexp{$MAPLOOP}, match={XXX}, replace={Acta Crystallogr. A, Found. Crystallogr.}]
			\step[fieldsource=\regexp{$MAPLOOP}, match={XXX}, replace={Acta Crystallogr. B, Struct. Sci.}]
			\step[fieldsource=\regexp{$MAPLOOP}, match={XXX}, replace={Acta Crystallogr. C, Cryst. Struct. Commun.}]
			\step[fieldsource=\regexp{$MAPLOOP}, match={XXX}, replace={Acta Electron. (France)}]
			\step[fieldsource=\regexp{$MAPLOOP}, match={XXX}, replace={Acta Cyberno}]
			\step[fieldsource=\regexp{$MAPLOOP}, match={XXX}, replace={Acta Electron. Sin. (China)}]
			\step[fieldsource=\regexp{$MAPLOOP}, match={XXX}, replace={Acta Geod. Geophys. Montan. Hung.}]
			\step[fieldsource=\regexp{$MAPLOOP}, match={XXX}, replace={Acta Geophys. Pol.}]
			\step[fieldsource=\regexp{$MAPLOOP}, match={XXX}, replace={Acta Geophys. Sin. (China)}]
			\step[fieldsource=\regexp{$MAPLOOP}, match={XXX}, replace={Acta Geophys. Sin. (USA)}]
			\step[fieldsource=\regexp{$MAPLOOP}, match={XXX}, replace={Acta Metall.}]
			\step[fieldsource=\regexp{$MAPLOOP}, match={XXX}, replace={Acta Mex. Cienc. Tecnol.}]
			\step[fieldsource=\regexp{$MAPLOOP}, match={XXX}, replace={Acta Phys. Hung.}]
			\step[fieldsource=\regexp{$MAPLOOP}, match={XXX}, replace={Acta Phys. Pol. A}]
			\step[fieldsource=\regexp{$MAPLOOP}, match={XXX}, replace={Acta Phys. Pol. B}]
			\step[fieldsource=\regexp{$MAPLOOP}, match={XXX}, replace={Acta Phys. Sin.}]
			\step[fieldsource=\regexp{$MAPLOOP}, match={XXX}, replace={Acta Phys. Slovaca}]
			\step[fieldsource=\regexp{$MAPLOOP}, match={XXX}, replace={Acta Politec. Mex.}]
			\step[fieldsource=\regexp{$MAPLOOP}, match={XXX}, replace={Acta Polytech. Scand. Appl. Phys. Ser.}]
			\step[fieldsource=\regexp{$MAPLOOP}, match={XXX}, replace={Acta Polytech. Scand. Chem. Technol.}]
			\step[fieldsource=\regexp{$MAPLOOP}, match={XXX}, replace={Metall. Ser.}]
			\step[fieldsource=\regexp{$MAPLOOP}, match={XXX}, replace={Acta Polytech. Scand. Electr. Eng. Ser.}]
			\step[fieldsource=\regexp{$MAPLOOP}, match={XXX}, replace={Acta Polytech. Scand. Math. Comput. Sci. Ser.}]
			\step[fieldsource=\regexp{$MAPLOOP}, match={XXX}, replace={Acta Polytech. Scand. Mech. Eng. Ser.}]
			\step[fieldsource=\regexp{$MAPLOOP}, match={XXX}, replace={Acta Seismol. Sin.}]
			\step[fieldsource=\regexp{$MAPLOOP}, match={XXX}, replace={Acta Tech. Acad. Sci. Hung.}]
			\step[fieldsource=\regexp{$MAPLOOP}, match={XXX}, replace={Acta Tech. CSAV}]
			\step[fieldsource=\regexp{$MAPLOOP}, match={XXX}, replace={Acustica}]
			\step[fieldsource=\regexp{$MAPLOOP}, match={XXX}, replace={(AEU) Arch. Elektr. Ubertragung}]
			\step[fieldsource=\regexp{$MAPLOOP}, match={XXX}, replace={Akust. Zh.}]
			\step[fieldsource=\regexp{$MAPLOOP}, match={XXX}, replace={Algorithmica}]
			\step[fieldsource=\regexp{$MAPLOOP}, match={XXX}, replace={Alta Freq.}]
			\step[fieldsource=\regexp{$MAPLOOP}, match={XXX}, replace={An. Acad. Bras. Cienc.}]
			\step[fieldsource=\regexp{$MAPLOOP}, match={XXX}, replace={An. Fis.}]
			\step[fieldsource=\regexp{$MAPLOOP}, match={XXX}, replace={An. Mec. Electr.}]
			\step[fieldsource=\regexp{$MAPLOOP}, match={XXX}, replace={Angew. Inform.}]
			\step[fieldsource=\regexp{$MAPLOOP}, match={XXX}, replace={Ann. Inst. Henri Poincare Phys. Theor.}]
			\step[fieldsource=\regexp{$MAPLOOP}, match={XXX}, replace={Ann. Soc. Sci. Brux. I, Sci. Math. Astron. Phys.}]
			\step[fieldsource=\regexp{$MAPLOOP}, match={XXX}, replace={Arch. Elektr. Uebertrag. (AEU)}] 
			\step[fieldsource=\regexp{$MAPLOOP}, match={XXX}, replace={Arch. Elektron. Uebertrag. Tech.}] 
			\step[fieldsource=\regexp{$MAPLOOP}, match={XXX}, replace={Arch. Elektrotech. (Poland)}]
			\step[fieldsource=\regexp{$MAPLOOP}, match={XXX}, replace={Arch. Elektrotech. (Germany)}]
			\step[fieldsource=\regexp{$MAPLOOP}, match={XXX}, replace={Ark. Fys. Semin. Trondheim}]
			\step[fieldsource=\regexp{$MAPLOOP}, match={XXX}, replace={Astrofizika}]
			\step[fieldsource=\regexp{$MAPLOOP}, match={XXX}, replace={Astron. Nachr. (Germany)}]
			\step[fieldsource=\regexp{$MAPLOOP}, match={XXX}, replace={Astron. Tidsskr.}]
			\step[fieldsource=\regexp{$MAPLOOP}, match={XXX}, replace={Astron. Vestn.}]
			\step[fieldsource=\regexp{$MAPLOOP}, match={XXX}, replace={Astron. Zh.}]
			\step[fieldsource=\regexp{$MAPLOOP}, match={XXX}, replace={Atomwirtsch.-Atomtech.}]
			\step[fieldsource=\regexp{$MAPLOOP}, match={XXX}, replace={Atti Accad. Sci. Ist. Bologna CI. Sci. Fis.}]
			\step[fieldsource=\regexp{$MAPLOOP}, match={XXX}, replace={Rend. XIII}]
			\step[fieldsource=\regexp{$MAPLOOP}, match={XXX}, replace={Atti Accad. Sci. Torino I, CI. Sci. Fis. Math. Nat.}]
			\step[fieldsource=\regexp{$MAPLOOP}, match={XXX}, replace={Autom. Strum.}]
			\step[fieldsource=\regexp{$MAPLOOP}, match={XXX}, replace={Autom. Tech. Prax.}]
			\step[fieldsource=\regexp{$MAPLOOP}, match={XXX}, replace={Automatica}]
			\step[fieldsource=\regexp{$MAPLOOP}, match={XXX}, replace={Automatie}]
			\step[fieldsource=\regexp{$MAPLOOP}, match={XXX}, replace={Automatika}]
			\step[fieldsource=\regexp{$MAPLOOP}, match={XXX}, replace={Automatisierungstechnik}]
			\step[fieldsource=\regexp{$MAPLOOP}, match={XXX}, replace={Automatizace}]
			\step[fieldsource=\regexp{$MAPLOOP}, match={XXX}, replace={Automedica}]
			\step[fieldsource=\regexp{$MAPLOOP}, match={XXX}, replace={Avtom. Telemekh.}]
			\step[fieldsource=\regexp{$MAPLOOP}, match={XXX}, replace={Avtom. Vychisl. Tekh.}]
			\step[fieldsource=\regexp{$MAPLOOP}, match={XXX}, replace={Avtomatika}]
			\step[fieldsource=\regexp{$MAPLOOP}, match={XXX}, replace={Avtometriya}]
			\step[fieldsource=\regexp{$MAPLOOP}, match={ATZelektronik worldwide}, replace={ATZelektron. worldw.}]
			\step[fieldsource=\regexp{$MAPLOOP}, match={XXX}, replace={Ber. Bunsenges. Phys. Chem.}]
			\step[fieldsource=\regexp{$MAPLOOP}, match={XXX}, replace={Biofizika}]
			\step[fieldsource=\regexp{$MAPLOOP}, match={XXX}, replace={Biometrika}]
			\step[fieldsource=\regexp{$MAPLOOP}, match={XXX}, replace={Boll. Geofis. Teor. Appl.}]
			\step[fieldsource=\regexp{$MAPLOOP}, match={XXX}, replace={Bull. Acad. Serbe Sci. Arts cl. Sci. Tech.}]
			\step[fieldsource=\regexp{$MAPLOOP}, match={XXX}, replace={Bull. Annu. Soc. Suisse Chronom. Lab. Suisse.}]
			\step[fieldsource=\regexp{$MAPLOOP}, match={XXX}, replace={Rech. Horlog.}]
			\step[fieldsource=\regexp{$MAPLOOP}, match={XXX}, replace={Bull. Cl. Sci. Acad. R. Belg.}]
			\step[fieldsource=\regexp{$MAPLOOP}, match={XXX}, replace={Bull. Dir. Etud. Rech. A}]
			\step[fieldsource=\regexp{$MAPLOOP}, match={XXX}, replace={Bull. Dir. Etud. Rech. B}]
			\step[fieldsource=\regexp{$MAPLOOP}, match={XXX}, replace={Bull. Dir. Etud. Rech. C}]
			\step[fieldsource=\regexp{$MAPLOOP}, match={XXX}, replace={Bull. Liaison Rech. Inform. Autom.}]
			\step[fieldsource=\regexp{$MAPLOOP}, match={XXX}, replace={Bur. Etud. Autom.}]
			\step[fieldsource=\regexp{$MAPLOOP}, match={XXX}, replace={CFI-Ceram. Forum Int.-Ber. Dtsch. Keram. Ges.}]
			\step[fieldsource=\regexp{$MAPLOOP}, match={XXX}, replace={Chem. Scr.}]
			\step[fieldsource=\regexp{$MAPLOOP}, match={XXX}, replace={Ciel Terre}]
			\step[fieldsource=\regexp{$MAPLOOP}, match={XXX}, replace={Cybernetica}]
			\step[fieldsource=\regexp{$MAPLOOP}, match={XXX}, replace={Deut. Hydrogr. Z.}]
			\step[fieldsource=\regexp{$MAPLOOP}, match={XXX}, replace={Dokl. Akad. Nauk SSSR}]
			\step[fieldsource=\regexp{$MAPLOOP}, match={XXX}, replace={Electroacoustique}]
			\step[fieldsource=\regexp{$MAPLOOP}, match={XXX}, replace={Electrochim. Acta}]
			\step[fieldsource=\regexp{$MAPLOOP}, match={XXX}, replace={Electrochim. Metal.}]
			\step[fieldsource=\regexp{$MAPLOOP}, match={XXX}, replace={Elektor Electron.}]
			\step[fieldsource=\regexp{$MAPLOOP}, match={XXX}, replace={Elektr. Bahnen}]
			\step[fieldsource=\regexp{$MAPLOOP}, match={XXX}, replace={Elektr. Energ.-Tech.}]
			\step[fieldsource=\regexp{$MAPLOOP}, match={XXX}, replace={Elektr. Masch.}]
			\step[fieldsource=\regexp{$MAPLOOP}, match={XXX}, replace={Elektr. Stn.}]
			\step[fieldsource=\regexp{$MAPLOOP}, match={XXX}, replace={Elektrichestvo}]
			\step[fieldsource=\regexp{$MAPLOOP}, match={XXX}, replace={Elektrie}]
			\step[fieldsource=\regexp{$MAPLOOP}, match={XXX}, replace={Elektrizitaetswirtschaft}]
			\step[fieldsource=\regexp{$MAPLOOP}, match={XXX}, replace={Elektro}]
			\step[fieldsource=\regexp{$MAPLOOP}, match={XXX}, replace={Elektro-Anz.}]
			\step[fieldsource=\regexp{$MAPLOOP}, match={XXX}, replace={Elektro-Jahr}]
			\step[fieldsource=\regexp{$MAPLOOP}, match={XXX}, replace={Elektrokhimiya}]
			\step[fieldsource=\regexp{$MAPLOOP}, match={XXX}, replace={Elektron}]
			\step[fieldsource=\regexp{$MAPLOOP}, match={XXX}, replace={Elektron Int.}]
			\step[fieldsource=\regexp{$MAPLOOP}, match={XXX}, replace={Elektron. Entwitkl.}]
			\step[fieldsource=\regexp{$MAPLOOP}, match={XXX}, replace={Elektron. Ind}]
			\step[fieldsource=\regexp{$MAPLOOP}, match={XXX}, replace={Elektron. J.}]
			\step[fieldsource=\regexp{$MAPLOOP}, match={XXX}, replace={Elektron. Prax.}]
			\step[fieldsource=\regexp{$MAPLOOP}, match={XXX}, replace={Elektron. Tekh.}]
			\step[fieldsource=\regexp{$MAPLOOP}, match={XXX}, replace={Elektronica}]
			\step[fieldsource=\regexp{$MAPLOOP}, match={XXX}, replace={Elektronik}]
			\step[fieldsource=\regexp{$MAPLOOP}, match={XXX}, replace={Elektronika}]
			\step[fieldsource=\regexp{$MAPLOOP}, match={XXX}, replace={Elektroniker}]
			\step[fieldsource=\regexp{$MAPLOOP}, match={XXX}, replace={Elektronikschau}]
			\step[fieldsource=\regexp{$MAPLOOP}, match={XXX}, replace={Elektrosvyaz}]
			\step[fieldsource=\regexp{$MAPLOOP}, match={XXX}, replace={Elektrotech. Cas.}]
			\step[fieldsource=\regexp{$MAPLOOP}, match={Elektrotechnik und Informationstechnik}, replace={Elektrotech. Inf. Tech.}]
			\step[fieldsource=\regexp{$MAPLOOP}, match={XXX}, replace={Elektrotech. Obz.}]
			\step[fieldsource=\regexp{$MAPLOOP}, match={XXX}, replace={Elektrotechniek}]
			\step[fieldsource=\regexp{$MAPLOOP}, match={XXX}, replace={Elektrotechnik (Czechoslovakia)}]
			\step[fieldsource=\regexp{$MAPLOOP}, match={XXX}, replace={Elektrotechnik (Switzerland)}]
			\step[fieldsource=\regexp{$MAPLOOP}, match={XXX}, replace={Elektrotechnik (Germany)}]
			\step[fieldsource=\regexp{$MAPLOOP}, match={XXX}, replace={Elektrotechnika}]
			\step[fieldsource=\regexp{$MAPLOOP}, match={XXX}, replace={Elektrotehnika, Zagreb}]
			\step[fieldsource=\regexp{$MAPLOOP}, match={XXX}, replace={Elektrotekhnika}]
			\step[fieldsource=\regexp{$MAPLOOP}, match={XXX}, replace={Elektroteknikeren}]
			\step[fieldsource=\regexp{$MAPLOOP}, match={XXX}, replace={Elektrowaerme Int. B.}]
			\step[fieldsource=\regexp{$MAPLOOP}, match={XXX}, replace={Elettrificazione}]
			\step[fieldsource=\regexp{$MAPLOOP}, match={XXX}, replace={Elettron. Oggi}]
			\step[fieldsource=\regexp{$MAPLOOP}, match={XXX}, replace={Elettron. Telecomun.}]
			\step[fieldsource=\regexp{$MAPLOOP}, match={XXX}, replace={Elettrotecnica}]
			\step[fieldsource=\regexp{$MAPLOOP}, match={XXX}, replace={Elek. Med Aktuell Elektron.}]
			\step[fieldsource=\regexp{$MAPLOOP}, match={XXX}, replace={Elteknik}]
			\step[fieldsource=\regexp{$MAPLOOP}, match={XXX}, replace={Energ. Atomtech.}]
			\step[fieldsource=\regexp{$MAPLOOP}, match={XXX}, replace={Energ. Elettr.}]
			\step[fieldsource=\regexp{$MAPLOOP}, match={XXX}, replace={Energetica}]
			\step[fieldsource=\regexp{$MAPLOOP}, match={XXX}, replace={Energetik}]
			\step[fieldsource=\regexp{$MAPLOOP}, match={XXX}, replace={Energetika}]
			\step[fieldsource=\regexp{$MAPLOOP}, match={XXX}, replace={Energetyka}]
			\step[fieldsource=\regexp{$MAPLOOP}, match={XXX}, replace={Energia Nuclear}]
			\step[fieldsource=\regexp{$MAPLOOP}, match={XXX}, replace={Energie Technik (Germany)}]
			\step[fieldsource=\regexp{$MAPLOOP}, match={XXX}, replace={Energie Technik (Switzerland)}]
			\step[fieldsource=\regexp{$MAPLOOP}, match={XXX}, replace={Entropie}]
			\step[fieldsource=\regexp{$MAPLOOP}, match={XXX}, replace={ETZ}]
			\step[fieldsource=\regexp{$MAPLOOP}, match={XXX}, replace={ETZ Arch.}]
			\step[fieldsource=\regexp{$MAPLOOP}, match={XXX}, replace={Feingeraetetechnik}]
			\step[fieldsource=\regexp{$MAPLOOP}, match={XXX}, replace={Feinw. Tech. Messtech.}]
			\step[fieldsource=\regexp{$MAPLOOP}, match={XXX}, replace={Fert. Tech. Betr.}]
			\step[fieldsource=\regexp{$MAPLOOP}, match={XXX}, replace={Fis. Tecnol.}]
			\step[fieldsource=\regexp{$MAPLOOP}, match={XXX}, replace={Fiz. Khim. Obrab. Mater.}]
			\step[fieldsource=\regexp{$MAPLOOP}, match={XXX}, replace={Fiz. Met. Metalloved}]
			\step[fieldsource=\regexp{$MAPLOOP}, match={XXX}, replace={Fiz. Nizk. Temp.}]
			\step[fieldsource=\regexp{$MAPLOOP}, match={XXX}, replace={Fiz. Plazmy}]
			\step[fieldsource=\regexp{$MAPLOOP}, match={XXX}, replace={Fiz. Tekh. Poluprovodn.}]
			\step[fieldsource=\regexp{$MAPLOOP}, match={XXX}, replace={Fiz. Tverd. Tela}]
			\step[fieldsource=\regexp{$MAPLOOP}, match={XXX}, replace={Fiz.-Khim. Mekh. Mater.}]
			\step[fieldsource=\regexp{$MAPLOOP}, match={XXX}, replace={Fizika}]
			\step[fieldsource=\regexp{$MAPLOOP}, match={XXX}, replace={Forsch.-Ber. Landes Nordrh.-Westfal.}]
			\step[fieldsource=\regexp{$MAPLOOP}, match={XXX}, replace={Frequenz}]
			\step[fieldsource=\regexp{$MAPLOOP}, match={XXX}, replace={Fys. Tidsskr.}]
			\step[fieldsource=\regexp{$MAPLOOP}, match={XXX}, replace={G. Fis.}]
			\step[fieldsource=\regexp{$MAPLOOP}, match={XXX}, replace={Geliotekhnika}]
			\step[fieldsource=\regexp{$MAPLOOP}, match={XXX}, replace={Geochim. Cosmochim. Acta}]
			\step[fieldsource=\regexp{$MAPLOOP}, match={XXX}, replace={Haerterei-Tech. Mitt.}]
			\step[fieldsource=\regexp{$MAPLOOP}, match={XXX}, replace={Helv. Chim. Acta}]
			\step[fieldsource=\regexp{$MAPLOOP}, match={XXX}, replace={Helv. Med. Acta}]
			\step[fieldsource=\regexp{$MAPLOOP}, match={XXX}, replace={Helv. Phys. Acta}]
			\step[fieldsource=\regexp{$MAPLOOP}, match={XXX}, replace={Hochfreq. Electroakust}]
			\step[fieldsource=\regexp{$MAPLOOP}, match={XXX}, replace={Hoppe-Seylers Z. Physiol. Chem.}]
			\step[fieldsource=\regexp{$MAPLOOP}, match={XXX}, replace={Inf. Elektron.}]
			\step[fieldsource=\regexp{$MAPLOOP}, match={XXX}, replace={Inf. Elettron.}]
			\step[fieldsource=\regexp{$MAPLOOP}, match={XXX}, replace={Inform. Forsch. Entwickl.}]
			\step[fieldsource=\regexp{$MAPLOOP}, match={XXX}, replace={Inform. Spektrum}]
			\step[fieldsource=\regexp{$MAPLOOP}, match={XXX}, replace={Inform.-Fachber.}]
			\step[fieldsource=\regexp{$MAPLOOP}, match={XXX}, replace={Informatie}]
			\step[fieldsource=\regexp{$MAPLOOP}, match={XXX}, replace={Informatik}]
			\step[fieldsource=\regexp{$MAPLOOP}, match={XXX}, replace={Informatologia Yugosl.}]
			\step[fieldsource=\regexp{$MAPLOOP}, match={XXX}, replace={Informatyka}]
			\step[fieldsource=\regexp{$MAPLOOP}, match={XXX}, replace={Infowelt}]
			\step[fieldsource=\regexp{$MAPLOOP}, match={XXX}, replace={Ing. Electr. Mec.}]
			\step[fieldsource=\regexp{$MAPLOOP}, match={XXX}, replace={Ing. Mec. Electr.}]
			\step[fieldsource=\regexp{$MAPLOOP}, match={XXX}, replace={Ing.-Arch.}]
			\step[fieldsource=\regexp{$MAPLOOP}, match={XXX}, replace={Inzh.-Fiz. Zh.}]
			\step[fieldsource=\regexp{$MAPLOOP}, match={XXX}, replace={Izmer. Tekh.}]
			\step[fieldsource=\regexp{$MAPLOOP}, match={XXX}, replace={Izv. Akad. Nauk Arm. SSR Ser. Tekh. Nauk}]
			\step[fieldsource=\regexp{$MAPLOOP}, match={XXX}, replace={Izv. Akad. Nauk SSSR Energ. Transp.}]
			\step[fieldsource=\regexp{$MAPLOOP}, match={XXX}, replace={Izv. Akad. Nauk SSSR Fiz. Atmos. Okeana}]
			\step[fieldsource=\regexp{$MAPLOOP}, match={XXX}, replace={Izv. Akad. Nauk SSSR Fiz. Zemli}]
			\step[fieldsource=\regexp{$MAPLOOP}, match={XXX}, replace={Izv. Akad. Nauk SSSR Ser. Fiz.}]
			\step[fieldsource=\regexp{$MAPLOOP}, match={XXX}, replace={Izv. Vyssh. Uchebn. Zaved. Elektromekh.}]
			\step[fieldsource=\regexp{$MAPLOOP}, match={XXX}, replace={Izv. Vyssh. Uchebn. Zaved. Radioelektron.}]
			\step[fieldsource=\regexp{$MAPLOOP}, match={XXX}, replace={Izv. Vyssh. Uchebn. Zaved. Radiofiz.}]
			\step[fieldsource=\regexp{$MAPLOOP}, match={XXX}, replace={J. Chim Phys. Phys.-Chim Biol.}]
			\step[fieldsource=\regexp{$MAPLOOP}, match={XXX}, replace={Kernenergie}]
			\step[fieldsource=\regexp{$MAPLOOP}, match={XXX}, replace={Kerntechnik}]
			\step[fieldsource=\regexp{$MAPLOOP}, match={XXX}, replace={Khim. Fiz.}]
			\step[fieldsource=\regexp{$MAPLOOP}, match={XXX}, replace={Kibern. Vychisl. Tekh.}]
			\step[fieldsource=\regexp{$MAPLOOP}, match={XXX}, replace={Kibernetika}]
			\step[fieldsource=\regexp{$MAPLOOP}, match={XXX}, replace={Kristallografiya}]
			\step[fieldsource=\regexp{$MAPLOOP}, match={XXX}, replace={Kvantovaya Elektron. Mosk.}]
			\step[fieldsource=\regexp{$MAPLOOP}, match={XXX}, replace={Kybernetes}]
			\step[fieldsource=\regexp{$MAPLOOP}, match={XXX}, replace={Kybernetika}]
			\step[fieldsource=\regexp{$MAPLOOP}, match={XXX}, replace={Med. Tek.}]
			\step[fieldsource=\regexp{$MAPLOOP}, match={XXX}, replace={Mekh. Avtom. Proizvod.}]
			\step[fieldsource=\regexp{$MAPLOOP}, match={XXX}, replace={Meres Autom.}]
			\step[fieldsource=\regexp{$MAPLOOP}, match={XXX}, replace={Mesures}]
			\step[fieldsource=\regexp{$MAPLOOP}, match={XXX}, replace={Metallofizika}]
			\step[fieldsource=\regexp{$MAPLOOP}, match={XXX}, replace={Metalloved. Term. Obrab. Met.}]
			\step[fieldsource=\regexp{$MAPLOOP}, match={XXX}, replace={Meteorol. Gidrol.}]
			\step[fieldsource=\regexp{$MAPLOOP}, match={XXX}, replace={Meteorol. Rundsch.}]
			\step[fieldsource=\regexp{$MAPLOOP}, match={XXX}, replace={Metrol. Apl.}]
			\step[fieldsource=\regexp{$MAPLOOP}, match={XXX}, replace={Metrologia}]
			\step[fieldsource=\regexp{$MAPLOOP}, match={XXX}, replace={Medel. Simul.}]
			\step[fieldsource=\regexp{$MAPLOOP}, match={XXX}, replace={Nachr. Dok.}]
			\step[fieldsource=\regexp{$MAPLOOP}, match={XXX}, replace={Nachr.tech. Elektron.}]
			\step[fieldsource=\regexp{$MAPLOOP}, match={XXX}, replace={Naturwissentchaften}]
			\step[fieldsource=\regexp{$MAPLOOP}, match={XXX}, replace={Neue Tech.}]
			\step[fieldsource=\regexp{$MAPLOOP}, match={XXX}, replace={Neue Tech. Buero}]
			\step[fieldsource=\regexp{$MAPLOOP}, match={XXX}, replace={Nukleonika}]
			\step[fieldsource=\regexp{$MAPLOOP}, match={XXX}, replace={Numer. Math.}]
			\step[fieldsource=\regexp{$MAPLOOP}, match={XXX}, replace={Nuovo Cimento A}]
			\step[fieldsource=\regexp{$MAPLOOP}, match={XXX}, replace={Nuovo Cimento B}]
			\step[fieldsource=\regexp{$MAPLOOP}, match={XXX}, replace={Nouvo Cimento C}]
			\step[fieldsource=\regexp{$MAPLOOP}, match={XXX}, replace={Nuovo Cimento D}]
			\step[fieldsource=\regexp{$MAPLOOP}, match={XXX}, replace={Okeanologiya}]
			\step[fieldsource=\regexp{$MAPLOOP}, match={XXX}, replace={Opt. Spektrosk.}]
			\step[fieldsource=\regexp{$MAPLOOP}, match={XXX}, replace={Opt.-Mekh. Prom.}]
			\step[fieldsource=\regexp{$MAPLOOP}, match={XXX}, replace={Optik}]
			\step[fieldsource=\regexp{$MAPLOOP}, match={XXX}, replace={Photogrammetria}]
			\step[fieldsource=\regexp{$MAPLOOP}, match={XXX}, replace={Photonics Spectra}]
			\step[fieldsource=\regexp{$MAPLOOP}, match={XXX}, replace={Pis’ma Astron. Zh. }]
			\step[fieldsource=\regexp{$MAPLOOP}, match={XXX}, replace={Pis’ma Zh. Eksp. Teor. Fiz. }]
			\step[fieldsource=\regexp{$MAPLOOP}, match={XXX}, replace={Pis’ma Zh. Tekh. Fiz. }]
			\step[fieldsource=\regexp{$MAPLOOP}, match={XXX}, replace={Poverkhn., Fiz. Khim. Mekh.}]
			\step[fieldsource=\regexp{$MAPLOOP}, match={XXX}, replace={Pr. Inst. Elektrotech.}]
			\step[fieldsource=\regexp{$MAPLOOP}, match={XXX}, replace={Prib. Sist. Upr.}]
			\step[fieldsource=\regexp{$MAPLOOP}, match={XXX}, replace={Prib. Tekh. Eksp.}]
			\step[fieldsource=\regexp{$MAPLOOP}, match={XXX}, replace={Prikl. Mat. Mekh.}]
			\step[fieldsource=\regexp{$MAPLOOP}, match={XXX}, replace={Prikl. Mekh.}]
			\step[fieldsource=\regexp{$MAPLOOP}, match={XXX}, replace={Probl. Kibern.}]
			\step[fieldsource=\regexp{$MAPLOOP}, match={XXX}, replace={Probl. Peredachi Inf.}]
			\step[fieldsource=\regexp{$MAPLOOP}, match={XXX}, replace={Proc. K. Ned. Akad. Wet. B, Palaeontol.}]
			\step[fieldsource=\regexp{$MAPLOOP}, match={XXX}, replace={Anthropol. }]
			\step[fieldsource=\regexp{$MAPLOOP}, match={XXX}, replace={Programmirovanie}]
			\step[fieldsource=\regexp{$MAPLOOP}, match={XXX}, replace={Prz. Elektrotech.}]
			\step[fieldsource=\regexp{$MAPLOOP}, match={XXX}, replace={Prz. Telekomun.}]
			\step[fieldsource=\regexp{$MAPLOOP}, match={XXX}, replace={PT/Elektrotech. Elektron.}]
			\step[fieldsource=\regexp{$MAPLOOP}, match={XXX}, replace={Radio Fernsehen Elektron.}]
			\step[fieldsource=\regexp{$MAPLOOP}, match={XXX}, replace={Radiotekh. Elektron.}]
			\step[fieldsource=\regexp{$MAPLOOP}, match={XXX}, replace={Radiotekhnika Mosk.}]
			\step[fieldsource=\regexp{$MAPLOOP}, match={XXX}, replace={Rev. Acad. Cienc. Zaragoza}]
			\step[fieldsource=\regexp{$MAPLOOP}, match={XXX}, replace={Rev. Electrotec. (Argentina)}]
			\step[fieldsource=\regexp{$MAPLOOP}, match={XXX}, replace={Rev. Electrotec. (Spain)}]
			\step[fieldsource=\regexp{$MAPLOOP}, match={XXX}, replace={Rev. Energ.}]
			\step[fieldsource=\regexp{$MAPLOOP}, match={XXX}, replace={Rev. Esp. Electron. Rev. Geofis.}]
			\step[fieldsource=\regexp{$MAPLOOP}, match={XXX}, replace={Ric. Autom.}]
			\step[fieldsource=\regexp{$MAPLOOP}, match={XXX}, replace={Ric. Spettrosc.}]
			\step[fieldsource=\regexp{$MAPLOOP}, match={XXX}, replace={Robotersysteme}]
			\step[fieldsource=\regexp{$MAPLOOP}, match={XXX}, replace={Rozpr. Electrotech.}]
			\step[fieldsource=\regexp{$MAPLOOP}, match={XXX}, replace={Sadhana}]
			\step[fieldsource=\regexp{$MAPLOOP}, match={XXX}, replace={Schweiz. Tech. Z.}]
			\step[fieldsource=\regexp{$MAPLOOP}, match={XXX}, replace={Scientia}]
			\step[fieldsource=\regexp{$MAPLOOP}, match={XXX}, replace={Siemens Forsch. Entwickl. Ber.}]
			\step[fieldsource=\regexp{$MAPLOOP}, match={XXX}, replace={Sist. Autom.}]
			\step[fieldsource=\regexp{$MAPLOOP}, match={XXX}, replace={Sitzungsber. Oester. Akad. Wiss. Math.-}]
			\step[fieldsource=\regexp{$MAPLOOP}, match={XXX}, replace={Naturwiss. Kl. Abt. II (Austria)}]
			\step[fieldsource=\regexp{$MAPLOOP}, match={XXX}, replace={Spectrochim. Acta A, Mol. Spectrosc.}]
			\step[fieldsource=\regexp{$MAPLOOP}, match={XXX}, replace={Spectrochim. Acta B, At. Spectrosc.}]
			\step[fieldsource=\regexp{$MAPLOOP}, match={XXX}, replace={Sprache Datenverarb.}]
			\step[fieldsource=\regexp{$MAPLOOP}, match={XXX}, replace={Stanki Instrum.}]
			\step[fieldsource=\regexp{$MAPLOOP}, match={XXX}, replace={Steklo Keram.}]
			\step[fieldsource=\regexp{$MAPLOOP}, match={XXX}, replace={Svetotekhnika  TE Int.}]
			\step[fieldsource=\regexp{$MAPLOOP}, match={XXX}, replace={Tech. Bull. Vevey}]
			\step[fieldsource=\regexp{$MAPLOOP}, match={XXX}, replace={Tech. Mitt. Krupp (Engl. Ed.)}]
			\step[fieldsource=\regexp{$MAPLOOP}, match={XXX}, replace={Tech. Mitt. PTT}]
			\step[fieldsource=\regexp{$MAPLOOP}, match={XXX}, replace={Tech. Mitt. RFZ}]
			\step[fieldsource=\regexp{$MAPLOOP}, match={XXX}, replace={Technica}]
			\step[fieldsource=\regexp{$MAPLOOP}, match={XXX}, replace={Tecnica}]
			\step[fieldsource=\regexp{$MAPLOOP}, match={XXX}, replace={Teh. Fiz.}]
			\step[fieldsource=\regexp{$MAPLOOP}, match={XXX}, replace={Tehnika}]
			\step[fieldsource=\regexp{$MAPLOOP}, match={XXX}, replace={Tekh. Elektrodin.}]
			\step[fieldsource=\regexp{$MAPLOOP}, match={XXX}, replace={Tekh. Kibern.}]
			\step[fieldsource=\regexp{$MAPLOOP}, match={XXX}, replace={Tekh. Kino Telev.}]
			\step[fieldsource=\regexp{$MAPLOOP}, match={XXX}, replace={Tekh. Misul}]
			\step[fieldsource=\regexp{$MAPLOOP}, match={XXX}, replace={Telekomunikacije}]
			\step[fieldsource=\regexp{$MAPLOOP}, match={XXX}, replace={Telektronikk}]
			\step[fieldsource=\regexp{$MAPLOOP}, match={XXX}, replace={Teleteknik}]
			\step[fieldsource=\regexp{$MAPLOOP}, match={XXX}, replace={Teor. Mat. Fiz.}]
			\step[fieldsource=\regexp{$MAPLOOP}, match={XXX}, replace={Teploeoergetika}]
			\step[fieldsource=\regexp{$MAPLOOP}, match={XXX}, replace={Teplofiz. Vvs. Temp.}]
			\step[fieldsource=\regexp{$MAPLOOP}, match={XXX}, replace={Tidskr. Dok.}]
			\step[fieldsource=\regexp{$MAPLOOP}, match={XXX}, replace={TN Nachr.}]
			\step[fieldsource=\regexp{$MAPLOOP}, match={XXX}, replace={Toute Electron.}]
			\step[fieldsource=\regexp{$MAPLOOP}, match={XXX}, replace={Tr. Inst. Teor. Astron.}]
			\step[fieldsource=\regexp{$MAPLOOP}, match={XXX}, replace={Ukr. Fiz. Zh.}]
			\step[fieldsource=\regexp{$MAPLOOP}, match={XXX}, replace={Usp. Fiz. Nauk}]
			\step[fieldsource=\regexp{$MAPLOOP}, match={XXX}, replace={Vak.-Tech.}]
			\step[fieldsource=\regexp{$MAPLOOP}, match={XXX}, replace={VDE Fachiber.}]
			\step[fieldsource=\regexp{$MAPLOOP}, match={XXX}, replace={VDI Z.}]
			\step[fieldsource=\regexp{$MAPLOOP}, match={XXX}, replace={Vestn. Mashinostr.}]
			\step[fieldsource=\regexp{$MAPLOOP}, match={XXX}, replace={Vestn. Mosk. Univ. 15, Vychisl. Mat. Kibern.}]
			\step[fieldsource=\regexp{$MAPLOOP}, match={XXX}, replace={Vestn. Mosk. Univ. 3, Fiz. Astron.}]
			\step[fieldsource=\regexp{$MAPLOOP}, match={XXX}, replace={Vesti Akad. Navuk BSSR Ser. Fiz. Energ. Navuk}]
			\step[fieldsource=\regexp{$MAPLOOP}, match={XXX}, replace={VGB Kraftwerkstech. (Ger. Ed.)}]
			\step[fieldsource=\regexp{$MAPLOOP}, match={XXX}, replace={Vide Couches Minces}]
			\step[fieldsource=\regexp{$MAPLOOP}, match={XXX}, replace={Vistas Astron.}]
			\step[fieldsource=\regexp{$MAPLOOP}, match={XXX}, replace={Vopr. At. Nauki Tekh. Ser., Fiz. Radiats.}]
			\step[fieldsource=\regexp{$MAPLOOP}, match={XXX}, replace={Povrezhdenii Radiats. Materialoved.}]
			\step[fieldsource=\regexp{$MAPLOOP}, match={XXX}, replace={Vopr. At. Nauki Tekh. Ser., Obshch. Yad. Fiz.}]
			\step[fieldsource=\regexp{$MAPLOOP}, match={XXX}, replace={Vuoto Sci. Tecnol.}]
			\step[fieldsource=\regexp{$MAPLOOP}, match={XXX}, replace={Wiss. Z. Friedrich-Schiller-Univ. Jena}]
			\step[fieldsource=\regexp{$MAPLOOP}, match={XXX}, replace={Nat.wiss. Reihe}]
			\step[fieldsource=\regexp{$MAPLOOP}, match={XXX}, replace={Wiss. Z. Karl-Marx-Univ. Leipz. Math.-}]
			\step[fieldsource=\regexp{$MAPLOOP}, match={XXX}, replace={Nat.wiss. Reihe}]
			\step[fieldsource=\regexp{$MAPLOOP}, match={XXX}, replace={Wiss. Z. Tech. Hochsch. Ilmenau}]
			\step[fieldsource=\regexp{$MAPLOOP}, match={XXX}, replace={Wiss. Z. Tech. Univ. Dresd.}]
			\step[fieldsource=\regexp{$MAPLOOP}, match={XXX}, replace={Wiss. Z. Tech. Univ. Karl-Marx-Stadt}]
			\step[fieldsource=\regexp{$MAPLOOP}, match={XXX}, replace={Yad. Fiz.}]
			\step[fieldsource=\regexp{$MAPLOOP}, match={XXX}, replace={Z. Angew. Math. Mech.}]
			\step[fieldsource=\regexp{$MAPLOOP}, match={XXX}, replace={Z. Angew. Math. Phys.}]
			\step[fieldsource=\regexp{$MAPLOOP}, match={XXX}, replace={Z. Met.kd.}]
			\step[fieldsource=\regexp{$MAPLOOP}, match={XXX}, replace={Z. Nat. Forsch. A, Phys. Phys. Chem. Kosmophys.}]
			\step[fieldsource=\regexp{$MAPLOOP}, match={XXX}, replace={Z. Oper. Res. A, Theor.}]
			\step[fieldsource=\regexp{$MAPLOOP}, match={XXX}, replace={Z. Oper. Res. B, Prax.}]
			\step[fieldsource=\regexp{$MAPLOOP}, match={XXX}, replace={Z. Phys.}]
			\step[fieldsource=\regexp{$MAPLOOP}, match={XXX}, replace={Z. Phys. A, At. Nuclei}]
			\step[fieldsource=\regexp{$MAPLOOP}, match={XXX}, replace={Z. Phys. B, Condens. Matter}]
			\step[fieldsource=\regexp{$MAPLOOP}, match={XXX}, replace={Z. Phys. C, Part Fields}]
			\step[fieldsource=\regexp{$MAPLOOP}, match={XXX}, replace={Z. Phys. Chem. Neue Folge}]
			\step[fieldsource=\regexp{$MAPLOOP}, match={XXX}, replace={Z. Phys. Chem., Leipz.}]
			\step[fieldsource=\regexp{$MAPLOOP}, match={XXX}, replace={Z. Phys. D, At. Mol. Clusters}]
			\step[fieldsource=\regexp{$MAPLOOP}, match={XXX}, replace={Zavod. Lab.}]
			\step[fieldsource=\regexp{$MAPLOOP}, match={XXX}, replace={Zh. Eksp. Teor. Fiz.}]
			\step[fieldsource=\regexp{$MAPLOOP}, match={XXX}, replace={Zh. Fiz. Khim.}]
			\step[fieldsource=\regexp{$MAPLOOP}, match={XXX}, replace={Zh. Prikl. Mekh. Tekh. Fiz.}]
			\step[fieldsource=\regexp{$MAPLOOP}, match={XXX}, replace={Zh. Prikl. Spektrosk.}]
			\step[fieldsource=\regexp{$MAPLOOP}, match={XXX}, replace={Zh. Tekh. Fiz.}]
			\step[fieldsource=\regexp{$MAPLOOP}, match={XXX}, replace={Zh. Vychisl. Mat. Mat. Fiz.}]
			\step[fieldsource=\regexp{$MAPLOOP}, match={XXX}, replace={Zisin, J. Seismol. Soc. Jpn.}] 
			\step[fieldsource=\regexp{$MAPLOOP}, match={Production}, replace={Prod.}]
			\step[fieldsource=\regexp{$MAPLOOP}, match={Reliability}, replace={Rel.}]
			\step[fieldsource=\regexp{$MAPLOOP}, match={Report}, replace={Rep.}]
			\step[fieldsource=\regexp{$MAPLOOP}, match={Semiconductor}, replace={Semicond.}]
			\step[fieldsource=\regexp{$MAPLOOP}, match={Research}, replace={Res.}]
			\step[fieldsource=\regexp{$MAPLOOP}, match={Sensing}, replace={Sens.}]
			\step[fieldsource=\regexp{$MAPLOOP}, match={Resonance}, replace={Reson.}]
			\step[fieldsource=\regexp{$MAPLOOP}, match={Series}, replace={Ser.}]
			\step[fieldsource=\regexp{$MAPLOOP}, match={Resources}, replace={Resour.}]
			\step[fieldsource=\regexp{$MAPLOOP}, match={Simulation}, replace={Simul.}]
			\step[fieldsource=\regexp{$MAPLOOP}, match={Reviews}, replace={Rev.}]
			\step[fieldsource=\regexp{$MAPLOOP}, match={Review}, replace={Rev.}]
			\step[fieldsource=\regexp{$MAPLOOP}, match={Singapore}, replace={Singap.}]
			\step[fieldsource=\regexp{$MAPLOOP}, match={Robotics}, replace={Robot.}]
			\step[fieldsource=\regexp{$MAPLOOP}, match={Sistema}, replace={Sist.}]
			\step[fieldsource=\regexp{$MAPLOOP}, match={Royal}, replace={Roy.}]
			\step[fieldsource=\regexp{$MAPLOOP}, match={Society}, replace={Soc.}]
			\step[fieldsource=\regexp{$MAPLOOP}, match={Safety}, replace={Saf.}]
			\step[fieldsource=\regexp{$MAPLOOP}, match={Sociological}, replace={Sociol.}]
			\step[fieldsource=\regexp{$MAPLOOP}, match={Satellite}, replace={Satell.}]
			\step[fieldsource=\regexp{$MAPLOOP}, match={Software}, replace={Softw.}]
			\step[fieldsource=\regexp{$MAPLOOP}, match={Scandinavian}, replace={Scand.}]
			\step[fieldsource=\regexp{$MAPLOOP}, match={Solar}, replace={Sol.}]
			\step[fieldsource=\regexp{$MAPLOOP}, match={Sciences}, replace={Sci.}]
			\step[fieldsource=\regexp{$MAPLOOP}, match={Science}, replace={Sci.}]
			\step[fieldsource=\regexp{$MAPLOOP}, match={Soviet}, replace={Sov.}]
			\step[fieldsource=\regexp{$MAPLOOP}, match={Section}, replace={Sect.}]
			\step[fieldsource=\regexp{$MAPLOOP}, match={Spectroscopy}, replace={Spectrosc.}]
			\step[fieldsource=\regexp{$MAPLOOP}, match={Security}, replace={Secur.}]
			\step[fieldsource=\regexp{$MAPLOOP}, match={Spectrum}, replace={Spectr.}]
			\step[fieldsource=\regexp{$MAPLOOP}, match={Seismology}, replace={Seismol.}]
			\step[fieldsource=\regexp{$MAPLOOP}, match={Speculations}, replace={Specul.}]
			\step[fieldsource=\regexp{$MAPLOOP}, match={Selected}, replace={Sel.}]
			\step[fieldsource=\regexp{$MAPLOOP}, match={Statistics}, replace={Statist.}]
			\step[fieldsource=\regexp{$MAPLOOP}, match={Structures}, replace={Struct.}]
			\step[fieldsource=\regexp{$MAPLOOP}, match={Structure}, replace={Struct.}]
			\step[fieldsource=\regexp{$MAPLOOP}, match={Terrestrial}, replace={Terr.}]
			\step[fieldsource=\regexp{$MAPLOOP}, match={Studies}, replace={Stud.}]
			\step[fieldsource=\regexp{$MAPLOOP}, match={Theoretical}, replace={Theor.}]
			\step[fieldsource=\regexp{$MAPLOOP}, match={Superconductivity}, replace={Supercond.}]
			\step[fieldsource=\regexp{$MAPLOOP}, match={Transactions}, replace={Trans.}]
			\step[fieldsource=\regexp{$MAPLOOP}, match={Supplement}, replace={Suppl.}]
			\step[fieldsource=\regexp{$MAPLOOP}, match={Translation}, replace={Transl.}]
			\step[fieldsource=\regexp{$MAPLOOP}, match={Surface}, replace={Surf.}]
			\step[fieldsource=\regexp{$MAPLOOP}, match={Transmission}, replace={Transmiss.}]
			\step[fieldsource=\regexp{$MAPLOOP}, match={Survey}, replace={Surv.}]
			\step[fieldsource=\regexp{$MAPLOOP}, match={Transportation}, replace={Transp.}]
			\step[fieldsource=\regexp{$MAPLOOP}, match={Sustainable}, replace={Sustain.}]
			\step[fieldsource=\regexp{$MAPLOOP}, match={Tutorials}, replace={Tut.}]
			\step[fieldsource=\regexp{$MAPLOOP}, match={Symposium}, replace={Symp.}]
			\step[fieldsource=\regexp{$MAPLOOP}, match={Ultrasonic}, replace={Ultrason.}]
			\step[fieldsource=\regexp{$MAPLOOP}, match={Systems}, replace={Syst.}]
			\step[fieldsource=\regexp{$MAPLOOP}, match={System}, replace={Syst.}]
			\step[fieldsource=\regexp{$MAPLOOP}, match={University}, replace={Univ.}]
			\step[fieldsource=\regexp{$MAPLOOP}, match={\detokenize{Universität}}, replace={Univ.}]
			\step[fieldsource=\regexp{$MAPLOOP}, match={\detokenize{Université}}, replace={Univ.}]
			\step[fieldsource=\regexp{$MAPLOOP}, match={Vacuum}, replace={Vac.}]
			\step[fieldsource=\regexp{$MAPLOOP}, match={Vehicles}, replace={Veh.}]
			\step[fieldsource=\regexp{$MAPLOOP}, match={Vehicle}, replace={Veh.}]
			\step[fieldsource=\regexp{$MAPLOOP}, match={Vehicular}, replace={Veh.}]
			\step[fieldsource=\regexp{$MAPLOOP}, match={Technology}, replace={Technol.}]
			\step[fieldsource=\regexp{$MAPLOOP}, match={Technological}, replace={Technol.}]
			\step[fieldsource=\regexp{$MAPLOOP}, match={Vibration}, replace={Vib.}]
			\step[fieldsource=\regexp{$MAPLOOP}, match={Telecommunications}, replace={Telecommun.}]
			\step[fieldsource=\regexp{$MAPLOOP}, match={Visual}, replace={Vis.}]
			\step[fieldsource=\regexp{$MAPLOOP}, match={Television}, replace={Telev.}]
			\step[fieldsource=\regexp{$MAPLOOP}, match={Welding}, replace={Weld.}]
			\step[fieldsource=\regexp{$MAPLOOP}, match={Temperature}, replace={Temp.}]
			\step[fieldsource=\regexp{$MAPLOOP}, match={Working}, replace={Work.}]
			\step[fieldsource=\regexp{$MAPLOOP}, match={Learning}, replace={Learn.}]
			\step[fieldsource=\regexp{$MAPLOOP}, match={Measurement}, replace={Meas.}]
			\step[fieldsource=\regexp{$MAPLOOP}, match={Letters}, replace={Lett.}]
			\step[fieldsource=\regexp{$MAPLOOP}, match={Letter}, replace={Lett.}]
			\step[fieldsource=\regexp{$MAPLOOP}, match={Mechanical}, replace={Mech.}]
			\step[fieldsource=\regexp{$MAPLOOP}, match={Mechanics}, replace={Mech.}]
			\step[fieldsource=\regexp{$MAPLOOP}, match={Mechanic}, replace={Mech.}]
			\step[fieldsource=\regexp{$MAPLOOP}, match={Lightwave}, replace={Lightw.}]
			\step[fieldsource=\regexp{$MAPLOOP}, match={Medical}, replace={Med.}]
			\step[fieldsource=\regexp{$MAPLOOP}, match={Logic, Logical}, replace={Log.}]
			\step[fieldsource=\regexp{$MAPLOOP}, match={Metals}, replace={Met.}]
			\step[fieldsource=\regexp{$MAPLOOP}, match={Luminescence}, replace={Lumin.}]
			\step[fieldsource=\regexp{$MAPLOOP}, match={Metallurgy}, replace={Metall.}]
			\step[fieldsource=\regexp{$MAPLOOP}, match={Machines}, replace={Mach.}]
			\step[fieldsource=\regexp{$MAPLOOP}, match={Machine}, replace={Mach.}]
			\step[fieldsource=\regexp{$MAPLOOP}, match={Meteorology}, replace={Meteorol.}]
			\step[fieldsource=\regexp{$MAPLOOP}, match={Magazine}, replace={Mag.}]
			\step[fieldsource=\regexp{$MAPLOOP}, match={Metropolitan}, replace={Metrop.}]
			\step[fieldsource=\regexp{$MAPLOOP}, match={Magnetics}, replace={Magn.}]
			\step[fieldsource=\regexp{$MAPLOOP}, match={Mexican, Mexico}, replace={Mex.}]
			\step[fieldsource=\regexp{$MAPLOOP}, match={Management}, replace={Manage.}]
			\step[fieldsource=\regexp{$MAPLOOP}, match={Microelectromechanical}, replace={Microelectromech.}]
			\step[fieldsource=\regexp{$MAPLOOP}, match={Managing}, replace={Manag.}]
			\step[fieldsource=\regexp{$MAPLOOP}, match={Microgravity}, replace={Microgr.}]
			\step[fieldsource=\regexp{$MAPLOOP}, match={Manufacturing}, replace={Manuf.}]
			\step[fieldsource=\regexp{$MAPLOOP}, match={Microscopy}, replace={Microsc.}]
			\step[fieldsource=\regexp{$MAPLOOP}, match={Marine}, replace={Mar.}]
			\step[fieldsource=\regexp{$MAPLOOP}, match={Microwaves}, replace={Microw.}]
			\step[fieldsource=\regexp{$MAPLOOP}, match={Microwave}, replace={Microw.}]
			\step[fieldsource=\regexp{$MAPLOOP}, match={Materials}, replace={Mater.}]
			\step[fieldsource=\regexp{$MAPLOOP}, match={Material}, replace={Mater.}]
			\step[fieldsource=\regexp{$MAPLOOP}, match={Military}, replace={Mil.}]
			\step[fieldsource=\regexp{$MAPLOOP}, match={Modeling}, replace={Model.}]
			\step[fieldsource=\regexp{$MAPLOOP}, match={Modelling}, replace={Model.}]
			\step[fieldsource=\regexp{$MAPLOOP}, match={Oceanic}, replace={Ocean.}]
			\step[fieldsource=\regexp{$MAPLOOP}, match={Molecular}, replace={Mol.}]
			\step[fieldsource=\regexp{$MAPLOOP}, match={Oceanography}, replace={Oceanogr.}]
			\step[fieldsource=\regexp{$MAPLOOP}, match={Monitoring}, replace={Monit.}]
			\step[fieldsource=\regexp{$MAPLOOP}, match={Occupation}, replace={Occupat.}]
			\step[fieldsource=\regexp{$MAPLOOP}, match={Multiphysics}, replace={Multiphys.}]
			\step[fieldsource=\regexp{$MAPLOOP}, match={Operational}, replace={Oper.}]
			\step[fieldsource=\regexp{$MAPLOOP}, match={Nanobioscience}, replace={Nanobiosci.}]
			\step[fieldsource=\regexp{$MAPLOOP}, match={Optical}, replace={Opt.}]
			\step[fieldsource=\regexp{$MAPLOOP}, match={Nanotechnology}, replace={Nanotechnol.}]
			\step[fieldsource=\regexp{$MAPLOOP}, match={Optics}, replace={Opt.}]
			\step[fieldsource=\regexp{$MAPLOOP}, match={National}, replace={Nat.}]
			\step[fieldsource=\regexp{$MAPLOOP}, match={Optimization}, replace={Optim.}]
			\step[fieldsource=\regexp{$MAPLOOP}, match={Naval}, replace={Nav.}]
			\step[fieldsource=\regexp{$MAPLOOP}, match={Organization}, replace={Org.}]
			\step[fieldsource=\regexp{$MAPLOOP}, match={Networking}, replace={Netw.}]
			\step[fieldsource=\regexp{$MAPLOOP}, match={Networked}, replace={Netw.}]
			\step[fieldsource=\regexp{$MAPLOOP}, match={Network}, replace={Netw.}]
			\step[fieldsource=\regexp{$MAPLOOP}, match={Packaging}, replace={Packag.}]
			\step[fieldsource=\regexp{$MAPLOOP}, match={Newsletter}, replace={Newslett.}]
			\step[fieldsource=\regexp{$MAPLOOP}, match={Particle}, replace={Part.}]
			\step[fieldsource=\regexp{$MAPLOOP}, match={Nondestructive}, replace={Nondestruct.}]
			\step[fieldsource=\regexp{$MAPLOOP}, match={Patent}, replace={Pat.}]
			\step[fieldsource=\regexp{$MAPLOOP}, match={Nuclear}, replace={Nucl.}]
			\step[fieldsource=\regexp{$MAPLOOP}, match={Performance}, replace={Perform.}]
			\step[fieldsource=\regexp{$MAPLOOP}, match={Numerical}, replace={Numer.}]
			\step[fieldsource=\regexp{$MAPLOOP}, match={Personal}, replace={Pers.}]
			\step[fieldsource=\regexp{$MAPLOOP}, match={Observations}, replace={Observ.}]
			\step[fieldsource=\regexp{$MAPLOOP}, match={Philosophical}, replace={Philos.}]
			\step[fieldsource=\regexp{$MAPLOOP}, match={Photonics}, replace={Photon.}]
			\step[fieldsource=\regexp{$MAPLOOP}, match={Productivity}, replace={Productiv.}]
			\step[fieldsource=\regexp{$MAPLOOP}, match={Photovoltaics}, replace={Photovolt.}]
			\step[fieldsource=\regexp{$MAPLOOP}, match={Programming}, replace={Program.}]
			\step[fieldsource=\regexp{$MAPLOOP}, match={Physics}, replace={Phys.}]
			\step[fieldsource=\regexp{$MAPLOOP}, match={Progress}, replace={Prog.}]
			\step[fieldsource=\regexp{$MAPLOOP}, match={Physiology}, replace={Physiol.}]
			\step[fieldsource=\regexp{$MAPLOOP}, match={Propagation}, replace={Propag.}]
			\step[fieldsource=\regexp{$MAPLOOP}, match={Planetary}, replace={Planet.}]
			\step[fieldsource=\regexp{$MAPLOOP}, match={Psychology}, replace={Psychol.}]
			\step[fieldsource=\regexp{$MAPLOOP}, match={Pneumatics}, replace={Pneum.}]
			\step[fieldsource=\regexp{$MAPLOOP}, match={Quality}, replace={Qual.}]
			\step[fieldsource=\regexp{$MAPLOOP}, match={Pollution}, replace={Pollut.}]
			\step[fieldsource=\regexp{$MAPLOOP}, match={Quarterly}, replace={Quart.}]
			\step[fieldsource=\regexp{$MAPLOOP}, match={Polymer}, replace={Polym.}]
			\step[fieldsource=\regexp{$MAPLOOP}, match={Radiation}, replace={Radiat.}]
			\step[fieldsource=\regexp{$MAPLOOP}, match={Polytechnic}, replace={Polytech.}]
			\step[fieldsource=\regexp{$MAPLOOP}, match={Radiology}, replace={Radiol.}]
			\step[fieldsource=\regexp{$MAPLOOP}, match={Practice}, replace={Pract.}]
			\step[fieldsource=\regexp{$MAPLOOP}, match={Reactor}, replace={React.}]
			\step[fieldsource=\regexp{$MAPLOOP}, match={Precision}, replace={Precis.}]
			\step[fieldsource=\regexp{$MAPLOOP}, match={Receivers}, replace={Receiv.}]
			\step[fieldsource=\regexp{$MAPLOOP}, match={Principles}, replace={Princ.}]
			\step[fieldsource=\regexp{$MAPLOOP}, match={Recognition}, replace={Recognit.}]
			\step[fieldsource=\regexp{$MAPLOOP}, match={Proceedings}, replace={Proc.}]
			\step[fieldsource=\regexp{$MAPLOOP}, match={Record}, replace={Rec.}]
			\step[fieldsource=\regexp{$MAPLOOP}, match={Processing}, replace={Process.}]
			\step[fieldsource=\regexp{$MAPLOOP}, match={Rehabilitation}, replace={Rehabil.}]
			\step[fieldsource=\regexp{$MAPLOOP}, match={Conversion}, replace={Convers.}]
			\step[fieldsource=\regexp{$MAPLOOP}, match={Digital}, replace={Digit.}]
			\step[fieldsource=\regexp{$MAPLOOP}, match={Convention}, replace={Conv.}]
			\step[fieldsource=\regexp{$MAPLOOP}, match={Disclosure}, replace={Discl.}]
			\step[fieldsource=\regexp{$MAPLOOP}, match={Correspondence}, replace={Corresp.}]
			\step[fieldsource=\regexp{$MAPLOOP}, match={Discussions}, replace={Discuss.}]
			\step[fieldsource=\regexp{$MAPLOOP}, match={Critical}, replace={Crit.}]
			\step[fieldsource=\regexp{$MAPLOOP}, match={Dissertations}, replace={Diss.}]
			\step[fieldsource=\regexp{$MAPLOOP}, match={Crystal}, replace={Cryst.}]
			\step[fieldsource=\regexp{$MAPLOOP}, match={Publicly}, replace={Pub.}]
			\step[fieldsource=\regexp{$MAPLOOP}, match={Available}, replace={Avail.}]
			\step[fieldsource=\regexp{$MAPLOOP}, match={Specification}, replace={Spec.}]
			\step[fieldsource=\regexp{$MAPLOOP}, match={Distributed}, replace={Distrib.}]
			\step[fieldsource=\regexp{$MAPLOOP}, match={Crystallography}, replace={Crystallogr.}]
			\step[fieldsource=\regexp{$MAPLOOP}, match={Dynamics}, replace={Dyn.}]
			\step[fieldsource=\regexp{$MAPLOOP}, match={Cybernetics}, replace={Cybern.}]
			\step[fieldsource=\regexp{$MAPLOOP}, match={Earthquake}, replace={Earthq.}]
			\step[fieldsource=\regexp{$MAPLOOP}, match={Decision}, replace={Decis.}]
			\step[fieldsource=\regexp{$MAPLOOP}, match={Economics}, replace={Econ.}]
			\step[fieldsource=\regexp{$MAPLOOP}, match={Economic}, replace={Econ.}]
			\step[fieldsource=\regexp{$MAPLOOP}, match={Economical}, replace={Econ.}]
			\step[fieldsource=\regexp{$MAPLOOP}, match={Edition}, replace={Ed.}]
			\step[fieldsource=\regexp{$MAPLOOP}, match={Evolutionary}, replace={Evol.}]
			\step[fieldsource=\regexp{$MAPLOOP}, match={Education}, replace={Educ.}]
			\step[fieldsource=\regexp{$MAPLOOP}, match={Exhibition}, replace={Exhib.}]
			\step[fieldsource=\regexp{$MAPLOOP}, match={Electrical}, replace={Elect.}]
			\step[fieldsource=\regexp{$MAPLOOP}, match={Electric}, replace={Elect.}]
			\step[fieldsource=\regexp{$MAPLOOP}, match={Experimental}, replace={Exp.}]
			\step[fieldsource=\regexp{$MAPLOOP}, match={Electrification}, replace={Electrific.}]
			\step[fieldsource=\regexp{$MAPLOOP}, match={Exploratory}, replace={Explor.}]
			\step[fieldsource=\regexp{$MAPLOOP}, match={Electromagnetic}, replace={Electromagn.}]
			\step[fieldsource=\regexp{$MAPLOOP}, match={Exposition}, replace={Expo.}]
			\step[fieldsource=\regexp{$MAPLOOP}, match={Electroacoustic}, replace={Electroacoust.}]
			\step[fieldsource=\regexp{$MAPLOOP}, match={Express}, replace={Express}]
			\step[fieldsource=\regexp{$MAPLOOP}, match={Electronics}, replace={Electron.}]
			\step[fieldsource=\regexp{$MAPLOOP}, match={Electronic}, replace={Electron.}]
			\step[fieldsource=\regexp{$MAPLOOP}, match={Fabrication}, replace={Fabr.}]
			\step[fieldsource=\regexp{$MAPLOOP}, match={Emerging}, replace={Emerg.}]
			\step[fieldsource=\regexp{$MAPLOOP}, match={Faculty}, replace={Fac.}]
			\step[fieldsource=\regexp{$MAPLOOP}, match={Engineering}, replace={Eng.}]
			\step[fieldsource=\regexp{$MAPLOOP}, match={Engineers}, replace={Eng.}]
			\step[fieldsource=\regexp{$MAPLOOP}, match={Engineer}, replace={Eng.}]
			\step[fieldsource=\regexp{$MAPLOOP}, match={Ferroelectrics}, replace={Ferroelect.}]
			\step[fieldsource=\regexp{$MAPLOOP}, match={Environment}, replace={Environ.}]
			\step[fieldsource=\regexp{$MAPLOOP}, match={Francais, French}, replace={Fr.}]
			\step[fieldsource=\regexp{$MAPLOOP}, match={Equations}, replace={Equ.}]
			\step[fieldsource=\regexp{$MAPLOOP}, match={Frequency}, replace={Freq.}]
			\step[fieldsource=\regexp{$MAPLOOP}, match={Equipment}, replace={Equip.}]
			\step[fieldsource=\regexp{$MAPLOOP}, match={Foundation}, replace={Found.}]
			\step[fieldsource=\regexp{$MAPLOOP}, match={Ergonomics}, replace={Ergonom.}]
			\step[fieldsource=\regexp{$MAPLOOP}, match={Fundamental}, replace={Fundam.}]
			\step[fieldsource=\regexp{$MAPLOOP}, match={European}, replace={Eur.}]
			\step[fieldsource=\regexp{$MAPLOOP}, match={Generation}, replace={Gener.}]
			\step[fieldsource=\regexp{$MAPLOOP}, match={Evaluation}, replace={Eval.}]
			\step[fieldsource=\regexp{$MAPLOOP}, match={Geology}, replace={Geol.}]
			\step[fieldsource=\regexp{$MAPLOOP}, match={Geophysics}, replace={Geophys.}]
			\step[fieldsource=\regexp{$MAPLOOP}, match={Innovations}, replace={Innov.}]
			\step[fieldsource=\regexp{$MAPLOOP}, match={Innovation}, replace={Innov.}]
			\step[fieldsource=\regexp{$MAPLOOP}, match={Geoscience}, replace={Geosci.}]
			\step[fieldsource=\regexp{$MAPLOOP}, match={Institute}, replace={Inst.}]
			\step[fieldsource=\regexp{$MAPLOOP}, match={Institution}, replace={Inst.}]
			\step[fieldsource=\regexp{$MAPLOOP}, match={Graphics}, replace={Graph.}]
			\step[fieldsource=\regexp{$MAPLOOP}, match={Instrument}, replace={Instrum.}]
			\step[fieldsource=\regexp{$MAPLOOP}, match={Guidance}, replace={Guid.}]
			\step[fieldsource=\regexp{$MAPLOOP}, match={Instrumentation}, replace={Instrum.}]
			\step[fieldsource=\regexp{$MAPLOOP}, match={Harmonics}, replace={Harmon.}]
			\step[fieldsource=\regexp{$MAPLOOP}, match={Harmonic}, replace={Harmon.}]
			\step[fieldsource=\regexp{$MAPLOOP}, match={Insulation}, replace={Insul.}]
			\step[fieldsource=\regexp{$MAPLOOP}, match={History}, replace={Hist.}]
			\step[fieldsource=\regexp{$MAPLOOP}, match={Integrated}, replace={Integr.}]
			\step[fieldsource=\regexp{$MAPLOOP}, match={Horizon}, replace={Horiz.}]
			\step[fieldsource=\regexp{$MAPLOOP}, match={Intelligence}, replace={Intell.}]
			\step[fieldsource=\regexp{$MAPLOOP}, match={Hungary}, replace={Hung.}]
			\step[fieldsource=\regexp{$MAPLOOP}, match={Hungarian}, replace={Hung.}]
			\step[fieldsource=\regexp{$MAPLOOP}, match={Intelligent}, replace={Intell.}]
			\step[fieldsource=\regexp{$MAPLOOP}, match={Hydraulics}, replace={Hydraul.}]
			\step[fieldsource=\regexp{$MAPLOOP}, match={Interactions}, replace={Interact.}]
			\step[fieldsource=\regexp{$MAPLOOP}, match={Hydrology}, replace={Hydrol.}]
			\step[fieldsource=\regexp{$MAPLOOP}, match={Internationales}, replace={Int.}]
			\step[fieldsource=\regexp{$MAPLOOP}, match={International}, replace={Int.}]
			\step[fieldsource=\regexp{$MAPLOOP}, match={Illuminating}, replace={Illum.}]
			\step[fieldsource=\regexp{$MAPLOOP}, match={Isotopes}, replace={Isot.}]
			\step[fieldsource=\regexp{$MAPLOOP}, match={Imaging}, replace={Imag.}]
			\step[fieldsource=\regexp{$MAPLOOP}, match={Israel}, replace={Isr.}]
			\step[fieldsource=\regexp{$MAPLOOP}, match={Industrial}, replace={Ind.}]
			\step[fieldsource=\regexp{$MAPLOOP}, match={Japan}, replace={Jpn.}]
			\step[fieldsource=\regexp{$MAPLOOP}, match={Information}, replace={Inf.}]
			\step[fieldsource=\regexp{$MAPLOOP}, match={Journal}, replace={J.}]
			\step[fieldsource=\regexp{$MAPLOOP}, match={Informatics}, replace={Inform.}]
			\step[fieldsource=\regexp{$MAPLOOP}, match={Knowledge}, replace={Knowl.}]
			\step[fieldsource=\regexp{$MAPLOOP}, match={Laboratory}, replace={Lab.}]
			\step[fieldsource=\regexp{$MAPLOOP}, match={Laboratories}, replace={Lab.}]
			\step[fieldsource=\regexp{$MAPLOOP}, match={Mathematical}, replace={Math.}]
			\step[fieldsource=\regexp{$MAPLOOP}, match={Language}, replace={Lang.}]
			\step[fieldsource=\regexp{$MAPLOOP}, match={Mathematics}, replace={Math.}]
			\step[fieldsource=\regexp{$MAPLOOP}, match={Abstracts}, replace={Abstr.}]
			\step[fieldsource=\regexp{$MAPLOOP}, match={Analysis}, replace={Anal.}]
			\step[fieldsource=\regexp{$MAPLOOP}, match={Academy}, replace={Acad.}]
			\step[fieldsource=\regexp{$MAPLOOP}, match={Annals}, replace={Ann.}]
			\step[fieldsource=\regexp{$MAPLOOP}, match={Accelerator}, replace={Accel.}]
			\step[fieldsource=\regexp{$MAPLOOP}, match={Annual}, replace={Annu.}]
			\step[fieldsource=\regexp{$MAPLOOP}, match={Acoustics}, replace={Acoust.}]
			\step[fieldsource=\regexp{$MAPLOOP}, match={Apparatus}, replace={App.}]
			\step[fieldsource=\regexp{$MAPLOOP}, match={Active}, replace={Act.}]
			\step[fieldsource=\regexp{$MAPLOOP}, match={Applications}, replace={Appl.}]
			\step[fieldsource=\regexp{$MAPLOOP}, match={Administration}, replace={Admin.}]
			\step[fieldsource=\regexp{$MAPLOOP}, match={Applied}, replace={Appl.}]
			\step[fieldsource=\regexp{$MAPLOOP}, match={Administrative}, replace={Administ.}]
			\step[fieldsource=\regexp{$MAPLOOP}, match={Approximate}, replace={Approx.}]
			\step[fieldsource=\regexp{$MAPLOOP}, match={Advanced}, replace={Adv.}]
			\step[fieldsource=\regexp{$MAPLOOP}, match={Advances}, replace={Adv.}]
			\step[fieldsource=\regexp{$MAPLOOP}, match={Archives}, replace={Arch.}]
			\step[fieldsource=\regexp{$MAPLOOP}, match={Archive}, replace={Arch.}]
			\step[fieldsource=\regexp{$MAPLOOP}, match={Aeronautics}, replace={Aeronaut.}]
			\step[fieldsource=\regexp{$MAPLOOP}, match={Artificial}, replace={Artif.}]
			\step[fieldsource=\regexp{$MAPLOOP}, match={Aerospace}, replace={Aerosp.}]
			\step[fieldsource=\regexp{$MAPLOOP}, match={Assembly}, replace={Assem.}]
			\step[fieldsource=\regexp{$MAPLOOP}, match={Affective}, replace={Affect.}]
			\step[fieldsource=\regexp{$MAPLOOP}, match={Association}, replace={Assoc.}]
			\step[fieldsource=\regexp{$MAPLOOP}, match={Africa}, replace={Afr.}]
			\step[fieldsource=\regexp{$MAPLOOP}, match={African}, replace={Afr.}]
			\step[fieldsource=\regexp{$MAPLOOP}, match={Astronomy}, replace={Astron.}]
			\step[fieldsource=\regexp{$MAPLOOP}, match={Aircraft}, replace={Aircr.}]
			\step[fieldsource=\regexp{$MAPLOOP}, match={Astronautics}, replace={Astronaut.}]
			\step[fieldsource=\regexp{$MAPLOOP}, match={Algebraic}, replace={Algebr.}]
			\step[fieldsource=\regexp{$MAPLOOP}, match={Astrophysics}, replace={Astrophys.}]
			\step[fieldsource=\regexp{$MAPLOOP}, match={American}, replace={Amer.}]
			\step[fieldsource=\regexp{$MAPLOOP}, match={Atmosphere}, replace={Atmos.}]
			\step[fieldsource=\regexp{$MAPLOOP}, match={Atomic}, replace={At.}]
			\step[fieldsource=\regexp{$MAPLOOP}, match={Atoms}, replace={At.}]
			\step[fieldsource=\regexp{$MAPLOOP}, match={Broadcasting}, replace={Broadcast.}]
			\step[fieldsource=\regexp{$MAPLOOP}, match={Australasian}, replace={Australas.}]
			\step[fieldsource=\regexp{$MAPLOOP}, match={Bulletin}, replace={Bull.}]
			\step[fieldsource=\regexp{$MAPLOOP}, match={Australia}, replace={Aust.}]
			\step[fieldsource=\regexp{$MAPLOOP}, match={Bureau}, replace={Bur.}]
			\step[fieldsource=\regexp{$MAPLOOP}, match={{Automatic }}, replace={Autom.}]
			\step[fieldsource=\regexp{$MAPLOOP}, match={Business}, replace={Bus.}]
			\step[fieldsource=\regexp{$MAPLOOP}, match={Automation}, replace={Automat.}]
			\step[fieldsource=\regexp{$MAPLOOP}, match={Canadian}, replace={Can.}]
			\step[fieldsource=\regexp{$MAPLOOP}, match={Automotive}, replace={Automot.}]
			\step[fieldsource=\regexp{$MAPLOOP}, match={Automobiles}, replace={Automob.}]
			\step[fieldsource=\regexp{$MAPLOOP}, match={Automobile}, replace={Automob.}]
			\step[fieldsource=\regexp{$MAPLOOP}, match={Ceramic}, replace={Ceram.}]
			\step[fieldsource=\regexp{$MAPLOOP}, match={Chemical}, replace={Chem.}]
			\step[fieldsource=\regexp{$MAPLOOP}, match={Behavioral}, replace={Behav.}]
			\step[fieldsource=\regexp{$MAPLOOP}, match={Behavior}, replace={Behav.}]
			\step[fieldsource=\regexp{$MAPLOOP}, match={Chinese}, replace={Chin.}]
			\step[fieldsource=\regexp{$MAPLOOP}, match={Belgian}, replace={Belg.}]
			\step[fieldsource=\regexp{$MAPLOOP}, match={Climatology}, replace={Climatol.}]
			\step[fieldsource=\regexp{$MAPLOOP}, match={Biochemical}, replace={Biochem.}]
			\step[fieldsource=\regexp{$MAPLOOP}, match={Clinical}, replace={Clin.}]
			\step[fieldsource=\regexp{$MAPLOOP}, match={Bioinformatics}, replace={Bioinf.}]
			\step[fieldsource=\regexp{$MAPLOOP}, match={Cognitive}, replace={Cogn.}]
			\step[fieldsource=\regexp{$MAPLOOP}, match={Biology, Biological}, replace={Biol.}]
			\step[fieldsource=\regexp{$MAPLOOP}, match={Colloquium}, replace={Colloq.}]
			\step[fieldsource=\regexp{$MAPLOOP}, match={Kolloquium}, replace={Kolloq.}]
			\step[fieldsource=\regexp{$MAPLOOP}, match={Biomedical}, replace={Biomed.}]
			\step[fieldsource=\regexp{$MAPLOOP}, match={Communications}, replace={Commun.}]
			\step[fieldsource=\regexp{$MAPLOOP}, match={Communication}, replace={Commun.}]
			\step[fieldsource=\regexp{$MAPLOOP}, match={Biophysics}, replace={Biophys.}]
			\step[fieldsource=\regexp{$MAPLOOP}, match={Compatibility}, replace={Compat.}]
			\step[fieldsource=\regexp{$MAPLOOP}, match={British}, replace={Brit.}]
			\step[fieldsource=\regexp{$MAPLOOP}, match={Components}, replace={Compon.}]
			\step[fieldsource=\regexp{$MAPLOOP}, match={Component}, replace={Compon.}]
			\step[fieldsource=\regexp{$MAPLOOP}, match={Computational}, replace={Comput.}]
			\step[fieldsource=\regexp{$MAPLOOP}, match={Delivery}, replace={Del.}]
			\step[fieldsource=\regexp{$MAPLOOP}, match={Computers}, replace={Comput.}]
			\step[fieldsource=\regexp{$MAPLOOP}, match={Computer}, replace={Comput.}]
			\step[fieldsource=\regexp{$MAPLOOP}, match={Department}, replace={Dept.}]
			\step[fieldsource=\regexp{$MAPLOOP}, match={Computing}, replace={Comput.}]
			\step[fieldsource=\regexp{$MAPLOOP}, match={Design}, replace={Des.}]
			\step[fieldsource=\regexp{$MAPLOOP}, match={Condensed}, replace={Condens.}]
			\step[fieldsource=\regexp{$MAPLOOP}, match={Detector}, replace={Detect.}]
			\step[fieldsource=\regexp{$MAPLOOP}, match={Conferences}, replace={Conf.}]
			\step[fieldsource=\regexp{$MAPLOOP}, match={Conference}, replace={Conf.}]
			\step[fieldsource=\regexp{$MAPLOOP}, match={Development}, replace={Develop.}]
			\step[fieldsource=\regexp{$MAPLOOP}, match={Congress}, replace={Congr.}]
			\step[fieldsource=\regexp{$MAPLOOP}, match={Differential}, replace={Differ.}]
			\step[fieldsource=\regexp{$MAPLOOP}, match={Consumer}, replace={Consum.}]
			\step[fieldsource=\regexp{$MAPLOOP}, match={Digest}, replace={Dig.}] 
			\step[fieldsource=\regexp{$MAPLOOP}, match={{ of the }}, replace={{ }}]
			\step[fieldsource=\regexp{$MAPLOOP}, match={{ of }}, replace={{ }}]
			\step[fieldsource=\regexp{$MAPLOOP}, match={{Of }}, replace={{ }}]
			\step[fieldsource=\regexp{$MAPLOOP}, match={{ on }}, replace={{ }}]
			\step[fieldsource=\regexp{$MAPLOOP}, match={{ On }}, replace={{ }}]
			\step[fieldsource=\regexp{$MAPLOOP}, match={{On }}, replace={{ }}]
			\step[fieldsource=\regexp{$MAPLOOP}, match={{ in }}, replace={{ }}]
			\step[fieldsource=\regexp{$MAPLOOP}, match=\regexp{\\\x{26}}, replace={{and}}] 
			\step[fieldsource=\regexp{$MAPLOOP}, match={{, and }}, replace={{, }}]
			\step[fieldsource=\regexp{$MAPLOOP}, match={{, And }}, replace={{, }}]
			\step[fieldsource=\regexp{$MAPLOOP}, match={{ and }}, replace={{ }}]
			\step[fieldsource=\regexp{$MAPLOOP}, match={{ And }}, replace={{ }}]
			\step[fieldsource=\regexp{$MAPLOOP}, match={{ In }}, replace={{ }}]
			\step[fieldsource=\regexp{$MAPLOOP}, match={{In }}, replace={{ }}]
			\step[fieldsource=\regexp{$MAPLOOP}, match={{ the }}, replace={{ }}] 
			\step[fieldsource=\regexp{$MAPLOOP}, match={{ The }}, replace={{ }}] 
			\step[fieldsource=\regexp{$MAPLOOP}, match={{The }}, replace={}] 
			\step[fieldsource=\regexp{$MAPLOOP}, match={{ for }}, replace={{ }}] 
			\step[fieldsource=\regexp{$MAPLOOP}, match={First}, replace={1st}]
			\step[fieldsource=\regexp{$MAPLOOP}, match={Second}, replace={2nd}]
			\step[fieldsource=\regexp{$MAPLOOP}, match={Third}, replace={3rd}]
			\step[fieldsource=\regexp{$MAPLOOP}, match={Fourth}, replace={4th}]
			\step[fieldsource=\regexp{$MAPLOOP}, match={Fifth}, replace={5th}]
			\step[fieldsource=\regexp{$MAPLOOP}, match={Sixth}, replace={6th}]
			\step[fieldsource=\regexp{$MAPLOOP}, match={Seventh}, replace={7th}]
			\step[fieldsource=\regexp{$MAPLOOP}, match={Eighth}, replace={8th}]
			\step[fieldsource=\regexp{$MAPLOOP}, match={Ninth}, replace={9th}]
			\step[fieldsource=\regexp{$MAPLOOP}, match={Tenth}, replace={10th}]
			\step[fieldsource=\regexp{$MAPLOOP}, match={Eleventh}, replace={11th}]
			\step[fieldsource=\regexp{$MAPLOOP}, match={Twelfth}, replace={12th}]
			\step[fieldsource=\regexp{$MAPLOOP}, match={Thirteenth}, replace={13th}]
			\step[fieldsource=\regexp{$MAPLOOP}, match={Fourteenth}, replace={14th}]
			\step[fieldsource=\regexp{$MAPLOOP}, match={Fifteenth}, replace={15th}]
			\step[fieldsource=\regexp{$MAPLOOP}, match={Sixteenth}, replace={16th}]
			\step[fieldsource=\regexp{$MAPLOOP}, match={Seventeenth}, replace={17th}]
			\step[fieldsource=\regexp{$MAPLOOP}, match={Eighteenth}, replace={18th}]
			\step[fieldsource=\regexp{$MAPLOOP}, match={Nineteenth}, replace={19th}]
			\step[fieldsource=\regexp{$MAPLOOP}, match={Twentieth}, replace={20th}]
		}
	}
}

\newcommand{\withcoverpage}{1}
\newcommand{\titlestring}{Managing Inherent Risk: On the Conceptualization of Risk in Defense Systems}

\ifthenelse{\withcoverpage=1}{
	\usepackage{hologo}
\usepackage{listings}
\lstset{breaklines,basicstyle=\small,columns=fullflexible,basicstyle=\ttfamily,language={[plain]TeX}}
}{}

\makeatletter
\newcommand{\linebreakand}{%
  \end{@IEEEauthorhalign}
  \hfill\mbox{}\par
  \mbox{}\hfill\begin{@IEEEauthorhalign}
}
\makeatother

\begin{document}

\ifthenelse{\withcoverpage=1}{
	\newcommand{\yearstring}{2026}
\newcommand{\conferencestring}{2026 IEEE International Workshop on Technologies for Defense and Security}
\newcommand{\addressstring}{Torino, Italy}
\newcommand{\authorstring}{Braun, Niklas and Brettin, Leon J and Loba, Marvin and Maurer, Markus}
\newcommand{\doistring}{10.XXXX/XXXX.XXXX.XXXXXXX}
\newcommand{\doiurlstring}{https://doi.org/\doistring}
\newcommand{\conferencedatestring}{November 4--6, 2026}

\newcommand{\ispublished}{0}
\newcommand{\withcopyrightnotice}{1}

\twocolumn[
  \begin{@twocolumnfalse}
    \ifthenelse{\ispublished=1}{
      \ifthenelse{\withcopyrightnotice=1}{
        \Huge {IEEE copyright notice} \\ \\
        \large {\copyright\ \yearstring\ IEEE. Personal use of this material is permitted. Permission from IEEE must be obtained for all other uses, in any current or future media, including reprinting/republishing this material for advertising or promotional purposes, creating new collective works, for resale or redistribution to servers or lists, or reuse of any copyrighted component of this work in other works.} \\ \\
      }{}

      {\Large Published in the proceedings of the \emph{\conferencestring}, \addressstring, \conferencedatestring.} \\ \\ 
    
      {\Large DOI: \href{\doiurlstring}{\doistring}} \\ \\
    }{
      \ifthenelse{\withcopyrightnotice=1}{
        \Huge {IEEE copyright notice} \\ \\
        \large {\copyright\ \yearstring\ IEEE. Personal use of this material is permitted. Permission from IEEE must be obtained for all other uses, in any current or future media, including reprinting/republishing this material for advertising or promotional purposes, creating new collective works, for resale or redistribution to servers or lists, or reuse of any copyrighted component of this work in other works.} \\ \\
      }{}

    }

    \vspace{2.0cm}

  Cite as:

  \vspace{0.2cm}
  \ifthenelse{\ispublished=1}
  {
    \noindent\fbox{%
      \parbox{\textwidth}{%
          \authorstring, ``\titlestring,''
    in \emph{\conferencestring}, \addressstring, \yearstring, pp. 2042--2049, doi: 10.1109/ITSC58415.2024.10919986
      }%
    }
  }
  {
      \noindent\fbox{%
      \parbox{\textwidth}{%
        \authorstring, ``\titlestring,'' accepted for publication at the \emph{\conferencestring}, \addressstring, \yearstring. Preprint available at https://arxiv.org/abs/2608.28093.
      }%
    }
  }
  \vspace{2cm}
  
  \end{@twocolumnfalse}
  ]
  
\noindent\begin{minipage}{\textwidth}
  
  \hologo{BibTeX}:

  \ifnum\ispublished=1
    \begin{lstlisting}[frame=single]
@inproceedings{braun_managing_2026,
  author={{Braun}, Niklas and {Brettin}, Leon J and {Loba}, Marvin and {Maurer}, Markus},
  booktitle={2026 {IEEE} {International} {Workshop} on {Technologies} for {Defense} and {Security} {({TECHDEFENSE})}},
  title={{Managing} {Inherent} {Risk}: {On} the {Conceptualization} of {Risk} in {Defense} {Systems}},
  address={Torino, Italy},
  year={2026},
  pages={XXXX-YYYY},
  doi={10.XXXX/XXXX.XXXX.XXXXXXX},
  publisher={IEEE}
      }
    \end{lstlisting}
  \else
    \begin{lstlisting}[frame=single]
@misc{braun_managing_2026,
  author={Braun, Niklas and Brettin, Leon J and Loba, Marvin and Maurer, Markus},
  title={Managing Inherent Risk: On the Conceptualization of Risk in Defense Systems},
  year={2026},
  note={Accepted for publication at the 2026 IEEE International Workshop on Technologies for Defense and Security (TECHDEFENSE), Torino, Italy.},
  arxiv={arXiv:2608.28093}
}
    \end{lstlisting}
  \fi
\end{minipage}

\newpage
}{}

\title{\titlestring%
}

\author{\IEEEauthorblockN{Niklas Braun}
\IEEEauthorblockA{\textit{Technische Universität Braunschweig} \\
\textit{Institut für Regelungstechnik}\\
Braunschweig, Germany \\
niklas.braun@tu-braunschweig.de}
\and
\IEEEauthorblockN{Leon J. Brettin}
\IEEEauthorblockA{\textit{Technische Universität Braunschweig} \\
\textit{Institut für Regelungstechnik}\\
Braunschweig, Germany \\
l.brettin@tu-braunschweig.de}
\linebreakand
\IEEEauthorblockN{Marvin Loba}
\IEEEauthorblockA{\textit{Technische Universität Braunschweig} \\
\textit{Institut für Regelungstechnik}\\
Braunschweig, Germany \\
m.loba@tu-braunschweig.de}
\and
\IEEEauthorblockN{Markus Maurer}
\IEEEauthorblockA{\textit{Technische Universität Braunschweig} \\
\textit{Institut für Regelungstechnik}\\
Braunschweig, Germany \\
markus.maurer@tu-braunschweig.de}
}

\maketitle

\begin{abstract}
Certain defense systems are, by nature, deployed in a civilian environment in order to serve a defensive function for that environment.
However, the risk involved with their deployment and operation poses a challenge for the public acceptance of these systems.
Compared to safety engineering for civilian systems, the risk constellation is quite different.
While reducing risk of safety-critical systems is generally desirable, defense systems are required to produce intended mission effects, which may include harm to an attacking target.
In both cases, not all risk can be eliminated.
The complex risk constellation of defense systems has implications for the systems' designs.
We compare the concepts of risk in existing standards from both the civilian and the military domains.
An extended constellation of risks needs to be considered, including risk caused by external threats to security.
By including this risk constellation in public communication about defense systems, we aim to stimulate a productive debate.
\end{abstract}


\section{Introduction}
\label{sec:intro}
Defense systems are often inherently dangerous, even if they serve a desirable defensive purpose.
For example, a ground-based air defense (GBAD) system uses missiles with kinetic force and explosive materials to defend against airborne targets that pose a threat.
The physical effects of such a system may result in unintended harm even in the absence of a malfunction.
In the case of a GBAD system, even if the defensive action is executed as intended, damage to property and potentially even to persons may occur due to debris, for instance.
The defensive purpose is often related to civilian objects such as government buildings or other critical infrastructure.
Thus, a GBAD system is likely to be positioned in an otherwise civilian environment.

As a consequence, the risks caused by defense systems in a civilian environment need to be reduced to a tolerable level.
Tolerability depends strongly on the deployment location and situation, e.g., peacetime versus wartime.
In a purely military context, performance is key; and certain residual risks can be accepted by a procuring authority~\cite{milstd882echange1}.
Existing safety assurance methods in the defense sector aim at achieving a corresponding level of risk.
Societal stakeholder expectations regarding safety and security are often implicit and complex, but relevant to achieving societal acceptance, which is influenced by the relation between utility and risk~\cite{lucke_Akzeptanz_1995}.
For systems that are novel in their deployment context, a safety-by-design approach is required that considers risks explicitly and incorporates stakeholder expectations adequately.
Where political and military decisions need to be backed up by the public, acceptance should be considered \emph{ex ante} by manufacturers and operators.
Otherwise, certain defense systems may be subject to unexpected public discourse.

Engineers face a similar challenge in the context of Automated Driving Systems.
Their novel nature within the socio-technical system of road traffic leads to a lack of experience and proven-in-use arguments regarding safety assurance.
Risk-based approaches from an operational-context point of view have been established in recent literature, and industrial players communicate their utility, e.g.,~\cite{favaroBuildingCredibleCase2023,nolte_Anmerkungen_2025a,nolte_Anmerkungen_2025}.
Context-sensitive risk management has been addressed in various engineering domains, including comparisons of military and civilian risk perspectives~\cite{stamatelatosProbabilisticRiskAssessment2011,millerComparisonMilitaryCivil1985}. This paper considers that perspective under the specific condition of coexistence between civilian and military structures.

Several research and standardization activities regarding defense system safety are ongoing.
With this paper, we contribute an automotive perspective to safety engineering in the defense sector and explicitly consider defense-specific kinds of risk as part of a conceptual framework.
In the spirit of scientific discourse, we warmly invite experts to respond to this paper.

After describing the publicly available baseline regarding risk concepts from both domains in Section~\ref{sec:background}, we conceptualize three categories of risk in the context of defense systems in Section~\ref{sec:risk_concept}.
An illustrative example is outlined in Section~\ref{sec:example}.
We conclude the paper in Section~\ref{sec:conclusion} and point out future research directions.

\section{Background}
\label{sec:background}
In engineering in general, and specifically in the defense and automotive domains, several standards and codes address the main concepts of \emph{safety} and \emph{risk}.
We analyze and compare the core concepts in the existing literature below.

\subsection{General Concepts of Safety and Risk}

In engineering, \emph{safety} is tied to \emph{risk} and \emph{harm}.
In this context, a consensus has emerged on the relation between these concepts.
ISO~Guide~51 on ``Safety aspects -- guidelines for their inclusion in standards'' is an established reference for cross-domain and domain-specific safety standards~\cite{iso26262-1:2018,iec61508-4:2010}.

\subsubsection{Safety}

In ISO~Guide~51, \emph{safety} is defined as ``freedom from risk [...] which is not tolerable''~\cite[clause 3.14]{isoiecguide51:2014}, where \emph{tolerable} is considered synonymous with \emph{acceptable}.
The level of \emph{tolerable risk} shall be ``accepted in a given context based on the current values of society''~\cite[clause 3.15]{isoiecguide51:2014}.
In our opinion, \emph{tolerable} is a less ambiguous term, as \emph{acceptability of risk} bears potential for confusion with \emph{(societal) acceptance}, as explained in Subsection~\ref{subsubsec:acceptance}.
With different wording, but equivalent intent, IEC~61508 defines \emph{safety} as ``freedom from unacceptable risk''~\cite[clause 3.1.11]{iec61508-4:2010}, which is adopted as ``absence of unreasonable risk'' by the domain-specific automotive standards ISO~26262~\cite[clause 3.132]{iso26262-1:2018} and ISO~21448~\cite[clause 3.25]{iso21448:2022}.
The qualification of a level of risk as \emph{(un)reasonable} follows an assessment informed by available risk-assessment guidance and should, ideally, represent the \emph{(un)acceptability} threshold~\cite{iso26262-1:2018}.
However, these two are not necessarily the same.

Aside from the engineering definitions of \emph{safety}, the term can be considered an ``open signifier''~\cite{nolte_Anmerkungen_2025a} (after~\cite{fleischer_Safety_2023}) that leaves its meaning open to different contextual fillings~\cite{levi-strauss_Introduction_1950}.
As such, safety integrates diverse viewpoints across stakeholder groups.
On the downside, this open understanding is a challenge for negotiating specific safety-related issues with said stakeholders.

\subsubsection{Risk and Harm}

The engineering definition of \emph{risk} is always tied to the occurrence of \emph{harm}: ISO~Guide~51 defines \emph{risk} as the ``combination of the probability of occurrence of harm [...] and the severity of that harm''~\cite[clause 3.9]{isoiecguide51:2014}.
The scope of \emph{harm} varies between standards, but ISO Guide 51 defines \emph{harm} as ``injury or damage to the health of people, or damage to property or the environment''~\cite[clause 3.1]{isoiecguide51:2014}.

\subsubsection{Risk Acceptance Criteria}

In civilian safety engineering, \emph{risk acceptance criteria} are required by established safety processes.
The basis for such risk acceptance criteria should be the aforementioned societal moral concepts.
However, there is a translation gap between these societal moral concepts and risk acceptance criteria that are operable within risk management.
Risk acceptance criteria might, for example, be unclear, incomplete, or unsuitable to evaluate the kind of risk that is being estimated during risk analysis.

\subsubsection{Acceptance}
\label{subsubsec:acceptance}

Achieving a risk level lower than some risk acceptance criterion does not imply reaching societal acceptance~\cite{otway_acceptable_1982}.
Instead, \emph{acceptance} is a complex phenomenon influenced by system properties (risk being one of them), the individual and societal perceptions of these properties, attitudes and opinions, and complex interactions between stakeholders~\cite{lucke_Akzeptanz_1995}.
Risk perception is shaped by individual and societal factors and cannot be reduced to technical risk metrics~\cite{slovicPerceptionRisk2011}.
\citeauthor{otway_acceptable_1982} conclude that a technical conceptualization of \emph{risk} is helpful for safety engineering, but has shortcomings in the light of societal acceptability issues, which are always in need of contextualization~\cite{otway_acceptable_1982}.
This makes communication with stakeholders an important part of risk management~\cite{covello_Risk_1986}.

\subsubsection{Aggregated and Event-Level Risk}

According to the definition referenced above, \emph{risk} is conceptualized as being countable and associated with one specific harm, given the parameters of risk (probability and severity of harm).
This allows for considering many risks separately, supporting a divide-and-conquer strategy for safety engineering, where risks are analyzed and treated at an event level.
In contrast, aggregated risk covers an extended temporal and spatial context for a system or multiple systems.

The aggregated risk notion can be derived intuitively from established risk acceptance criteria and is required for an open context, where it is impossible to foresee and treat all event-level risks separately.
However, as \citeauthor{koopman_Breaking_2023} point out, managing risk purely on an aggregated level leaves potential for ``risk subsidy''~\cite{koopman_Breaking_2023}, where, despite achieving an overall positive net risk balance, a system might still show unacceptable behavior under specific conditions.

\subsection{Risk Management in Defense Engineering}

For the purposes of this paper, the considered weapon systems are designed to defend against external threats, i.e., to contribute to security.
We acknowledge that there is an open debate about the definability of defensive (or offensive, respectively) weapon systems.

On safety in defense systems, the US Department of Defense standard practice \mbox{MIL-STD-882E} Change~1 provides guidance on hazard identification as well as risk assessment and treatment throughout the system life cycle~\cite{milstd882echange1}.
According to this standard practice, the risk parameters are identified based on the context.
The risk remaining after risk reduction needs to be accepted by the procuring authority.
Additional guidance is provided by \mbox{AOP-15}, which describes risk assessment methods for non-nuclear munitions~\cite{aop15ed3}; our focus, however, is system safety.
The ALARP principle is central to the risk-management approach of DefStan~00-56~\cite{rangesafetygroupriskcommittee_Common_2020,defstan00056-1}.

In both standards, the definitions of safety are absolute, i.e., referring to ``[f]reedom from those conditions that can cause death, injury, occupational illness, damage to or loss of equipment or property, or damage to the environment''~\cite[clause 2.1.6]{aop15ed3}.
Although these definitions are absolute, the standards apply tolerable-risk practices in their risk management approaches.

Furthermore, it becomes clear from the standards that \emph{safety} refers to \emph{mishaps}, i.e., harm in the sense of the above definitions that is not intended by the designers and operators.
This is  relevant for defense systems, since the stakeholder groups affected by the risks caused by these systems are potentially different from the respective decision-makers.

\subsection{Risk Management in Automotive Engineering}

Weapon systems are subject to an inherently open and uncertain operating context. A comparable challenge exists in automotive systems engineering with the open-world nature of road traffic.
This, in combination with the complexity of automotive systems, leads to necessarily incomplete requirements, verification, and validation.
In particular, this finding is true for innovative systems like Automated Driving Systems\footnote{Automated Driving System (ADS): ``The hardware and software that are collectively capable of performing the entire [dynamic driving task] on a sustained basis [...]''~\cite[clause 3.2]{isosaepas22736:2021}}.
As a result, there is an ``inherent risk''~\cite{nolte_Anmerkungen_2025a} (after~\cite{maurer_inharente_2023}) that can be treated, but never eliminated.
This acknowledgment contributed to overcoming unrealistic expectations such as \emph{vision zero}.
The lack of conceptual clarity regarding safety in current ADS regulation remains an obstacle to successful market introduction~\cite{nolte_Conceptualizations_2026}.

Major research efforts have addressed managing risks in order to achieve safe systems despite these challenges.
For instance, an explicit representation of risk based on the operational context supports defining adequate requirements at an early design stage~\cite{salem_Risk_2024}.
On various abstraction levels, scenarios provide a useful means of communication regarding risk between stakeholders~\cite{ulbrich_Defining_2015}.

\subsection{Security and Cybersecurity}

Defense systems are supposed to contribute to security.
We distinguish cybersecurity as a specific domain and use \emph{security} in a broader sense for protection against threats to states, people, goods, and infrastructure.
\citeauthor{craigen_Defining_2014} attest the lack of a consensual definition of \emph{cybersecurity} and propose the definition as ``the organization and collection of resources, processes, and structures used to protect cyberspace and cyberspace-enabled systems from occurrences that misalign de jure from de facto property rights''~\cite{craigen_Defining_2014}.
ISO/SAE~21434 provides requirements to cybersecurity engineering in the automotive domain~\cite{isosae21434}.

In the domain of political science, \emph{security} is an open concept, similarly to \emph{safety}.
\citeauthor{romm_Defining_1993} criticizes the overuse and diverse use of the term \emph{national security}~\cite{romm_Defining_1993}.
\citeauthor{paleri_National_2022} understands \emph{security} as the (partial) absence of threats to a state and to its ability to defend against those threats which are directed at the nation's constituents (e.g., territory, people, and goods)~\cite{paleri_National_2022}.
However, the author acknowledges the lack of a distinctive definition likewise.

We refer to \emph{security} as the protection against threats in the sense of \citeauthor{paleri_National_2022}, being a desired emerging property of defense systems.
By no means do we consider the terminological work on this complete.

\subsection{Summary of Related Work}

Comparing the domains, the definitions of risk are compatible as well as the general notion of safety, despite the partially inconsistent use of the latter term.
However, what phenomena conceptually belong to risk and how a tolerable level of risk is to be achieved vary between the domains.
Still, the inherent risk is a common theme.
In the following section, we propose an automotive-driven perspective that should support acceptance of defense systems with inherent risk.

\section{Risk Conceptualization}
\label{sec:risk_concept}
As any technical system, defense systems require their developers to take design decisions in the light of design conflicts.
Development frameworks that explicitly incorporate stakeholder values and communicability have the potential to increase the acceptance of defense systems.
Managing risks plays a crucial role for
(1) resolving design conflicts in an ethically justifiable manner and
(2) reducing ethical and legal uncertainty for those who take deployment decisions.
In the following, we propose how risks can be structured, supporting the explicit balancing of risks which, in turn, should influence societal acceptability.

\subsection{Cause and Context of Risk}

In order to define the risk categories, we conceptualize that \emph{risk} is \emph{caused} by an entity in a certain \emph{context}.

In the field of safety engineering, innovative systems are considered to introduce risk to a given operating context~\cite{leveson_Engineering_2011}.
An example from the automotive domain is the market introduction of Automated Driving Systems in the context of road traffic.
Notwithstanding their potential to contribute to road safety, Automated Driving Systems change the sources, causes, and manifestations of risk in road traffic and become part of the overall socio-technical system road traffic, which in itself exhibits risks.
Therefore, the risk management process becomes a task of designing the Automated Driving Systems appropriately.
The operating context is considered given, therefore assigning the responsibility of achieving an acceptable level of risk (i.e., safety) to Automated Driving System developers.

The assignment of cause and context becomes different in the cases of malicious threats to newly introduced systems.
For example, cyber attacks on connected systems can lead to hazardous behavior of these systems, which can cause harm and therefore increase the risk.
This risk cannot be attributed to the system.
Nevertheless, since the system is part of the context in which the risk is caused, protection against such threats is demanded from the developers in order to achieve an acceptable level of risk in the light of threats.
Therefore, we consider it mandatory to include malicious action, both physically and digitally, in safety engineering.

For a defense system, risks linked to unintended harm can be dealt with similarly to civilian systems, even though the role of risk acceptance criteria varies between respective relevant standards~\cite{milstd882echange1,iec61508-4:2010}.
These risks are caused by the system and occur in their respective operating contexts.
Following the automotive line of argumentation, it is the developers' responsibility to achieve a tolerable level of these risks.

This assignment of responsibility is not as clear in the case of external threats in the sense of security.
An attacker may be considered the cause of risk since, were it not for an attacker, the risks due to a potential attack and the secondary risks caused by systems required to defend against, such attacks would not exist.
On the other hand, the presence of an attacker may be considered as context for the secondary risks, since these risks are partially controllable by the system designers and therefore caused by the system and its design.
By controllability, we refer to the idea that there exist design choices that lead to a desired level of risk in the final product.

We observe a distinction between the cause of risk (e.g., the system) and the context in which the risk occurs (e.g., the socio-technical environment).
Yet, this thought model is not without limitations.
Firstly, the distinction between cause and context requires a clear definition of the system boundary.
Given the increasingly connected and distributed nature of state-of-the-art defense systems, this boundary is hard to define, although required for system design anyway.
Secondly, cause and context are not trivial to assign as explained above.
The assignment depends on the perspectives of stakeholders and may even reach into the landscape of political debate.
Furthermore, the assignment of responsibility for risk management is linked, but not identical, to the assignment of liability and legal accountability for risk management.
The legal aspects, however, are not further addressed in this paper.

Despite these limitations we find the thought model of cause and context of risk helpful in order to distinguish three categories of risk in the context of defense systems which will be defined in the following section.
Furthermore, the model clarifies that not all risk is entirely controllable, supporting the communication of the inherent risk.

\subsection{Risk Categories}

We propose to explicitly distinguish three categories of risk for defense systems:

\begin{enumerate}
  \item risk caused by the system and linked to unintended harm,
  \item risk caused by the system and linked to intended harm, i.e., harm to an object that poses a threat, and
  \item risk caused by external threats, i.e., harm by an object that poses a threat.
\end{enumerate}

With this viewpoint, the unique risk constellations with defense systems become clear.
Reducing risk is subject to design trade-offs regarding other emerging properties, but still generally desirable.
While there is a bound to the achievable risk reduction~\cite{iec61508-1:2010,fowler_IEC_2022}, civilian safety engineering assumes \emph{the lower the risk, the better}.

For defense systems, managing risk of the first category follows a similar notion.
The safety of military personnel and workers falls under this category.
For example, the UK Defense Standard \mbox{00-56} Part~1 \cite{defstan00056-1} is partially motivated by the government's obligation to take care of those who are in military service.
In Western societies, we suspect an increasing awareness and societal expectation that military personnel are not exposed to unreasonable levels of risk, particularly at times of peace or during exercise.
Collateral damage also falls under the first category of risk.
For example, when a GBAD system defends against an airborne target, debris will inevitably fall to the ground and potentially cause damage to property or, depending on the situation, to the health of people.

The second category is specific to defense: Strictly following the definitions of civilian safety engineering standards, even risks linked to intended harm caused by the system are to be reduced.
However, strictly applying this idea would deprive the entire defense system of its purpose.
Therefore, harm that belongs to the intended mission effects needs to be excluded from risk management in the civilian sense.
Instead, the second category of risk should be managed through military decisions and requirements that are in line with \emph{jus in bello} which governs the conduct of armed conflict~\cite{lieblich_Facilitative_2019}.

The third category of risk includes potential harm resulting from an (undefended) external threat.
Cyber threats are treated separately and are not included in category three, because they are not the type of external threat that the GBAD system is intended to counter.
These threats are to be defended against by the defense systems considered in this paper.
Risk of the third category is caused by, for example, other military forces on the one hand, but may be influenced by the (suspected) existence of the defense system under development on the other hand.
A drastic example of this link is the arms race which, despite valid criticism, can be rationally explained as a result of the stakeholder need of (perceived) security.
This category is required in order to formulate an operable risk balance.
Without an external threat, there is little to no justification for introducing category one and two risks.
However, if risks due to external threats can be reduced sufficiently by a defense system, the overall risk balance may justify or even mandate the introduction of the defense system.

One key challenge with the third category of risk is that these risks are extremely hard to calculate.
External threats are only partially observable and rely on predictions about the future behavior of external actors.
As a result, the risk in this category is necessarily uncertain and based on assumptions; nevertheless, it is necessary to consider it in a risk balance.

\begin{table*}[!t]
\centering
\small
\caption{Three risk categories illustrated with a GBAD system.}
\label{tab:gbad-risk-categories}
\renewcommand{\arraystretch}{1.15}
\resizebox{\textwidth}{!}{%
\begin{tabular}{p{0.01\textwidth} p{0.08\textwidth} | p{0.19\textwidth} p{0.19\textwidth} | p{0.25\textwidth} p{0.25\textwidth}}
\hline
& \textbf{Category} & \textbf{Civilian view} & \textbf{Military view} & \textbf{Deployed GBAD} & \textbf{Non-deployed GBAD} \\
\hline
1 & Unintended harm & Unintended harm by the GBAD system is associated with risk and should be reduced. & Mishaps remain subject to safety risk management. & Debris from an interception or an uncommanded launch may harm civilians or property. & Transport, storage, or maintenance accidents may still cause unintended harm. \\
2 & Intended mission effects & Intended harm by the GBAD system is counted as harm under a strict civilian definition of risk. & The desired mission effect is not a safety risk; it is governed by military decisions and \emph{jus in bello}. & The missile destroys the airborne target as intended. & No intended mission effect occurs because the system does not engage a target. \\
3 & External threats & The external threat by the (undefended) target remains a source of risk in the operating context. & The system modifies the external-threat risk, so deployment must be compared with non-deployment. & GBAD can reduce the airborne threat, but the deployed system can itself become a target and affect escalation. & The airborne threat remains unmitigated, while the additional risk of deploying the GBAD system is absent. \\
\hline
\end{tabular}
}
\end{table*}

\subsection{Risk Acceptance Criteria}

Risk acceptance criteria are of particular importance for developing innovative systems with which there is little in-field experience regarding risk perception and acceptance.
The translation of societal expectations into risk acceptance criteria is a challenge that has not been fully resolved.
Challenges include inconsistent attitudes, complex constellations of risk perception, context dependence of acceptability~\cite{lucke_Akzeptanz_1995}, and the varying risk perception before versus after an incident.
Nevertheless, risk acceptance criteria as engineering artifacts contribute to the communication with and among engineers.

They are relevant to decision-makers who want or need to consider risk, such as procuring authorities, even where mission effectiveness may take priority.

This civilian approach to risk acceptance criteria differs from the military approach prescribed by the standards.
The military system safety standard \cite{milstd882echange1} requires the residual risk to be accepted by the procuring authority.
It is manufacturers' responsibility to design systems that an authority will accept, ultimately.
This highlights the need for a proactive design approach.

Quantitative approaches to risk acceptance criteria provide a link between empirical data of other, but similar, systems and safety requirements for novel systems.
Subsequently, these criteria can be decomposed and allocated across the system architecture and operational conditions.
Quantitative risk acceptance criteria set out with a basic principle of measuring the level of risk.
The principle is then parametrized with a quantitative reference.
However, mapping risk acceptance criteria to event-level risks is not trivial~\cite{koopman_Breaking_2023}.

Risk acceptance criteria rely on the engineering definition of risk, i.e., the probability of occurrence of harm and the severity of that harm.
Therefore, the context in which harm occurs as well as the causal chain of events leading to that harm are omitted.
Yet, the literature indicates that these aspects are important for individual and societal perceptions of risk; and in the context of defense systems, we suspect the influence to be especially drastic.

\subsection{Balancing Risks for Safety and Security}

Designing and operating a defense system are matters of balancing risks.
We propose to consider the three categories of risk when taking design and deployment decisions.
Safety in the broader sense (i.e., not in the technical sense as in absence of unreasonable risk) is an emerging property of a defense system expected by various stakeholders.
Similarly, the whole purpose of that system is to contribute to security.
Firstly, the system is supposed to defend against external threats.
Secondly, effective deterrence contributes to security as it changes an aggressor's cost balance of an attack, reducing the level of risk in the third category.
In terms of societal expectations, the overall risk balance should address safety and security needs.

Both objectives can be conflictive when it comes to the performance capability of the defense system and its safety assurance which, respectively, influence risks of categories one and three.
This conflict is intuitive under the consideration of limited design resources such as mass of a moving object or volume requirements in a kinetic unit.
This comparison does not constitute a risk subsidy in the sense of an unfair redistribution of risk across scenarios or groups of affected actors. The benefit and the collateral damage are assessed within the same scenario and within a closed risk balance, rather than aggregated across an entire operating domain.

\subsection{Risk Communication}
In states where deployment decisions of defense systems are dependent on political decisions backed up by society, a partially transparent, i.e. ``honest''~\cite[definition of \emph{transparency}, p.~22]{isoiecieee24748-7000:2022} communication of the risk balance may be necessary to achieve security through proactive expectation management.
We deem the risk categories helpful for stimulating a differentiated debate.
However, full transparency is not desirable since disclosing deliberate amounts of information publicly might increase the level of threat, deteriorating the overall risk balance.
This conflict should be resolved by identifying adequate concepts and language for stakeholder communication, which is future work.

\section{Example}
\label{sec:example}

The conceptualization of risk is illustrated using the example of a ground-based air defense (GBAD) system.
The system is intended to defend against attacks by airborne targets by launching a missile with a kinetic effector capable of destroying the target.

The overview in Table~\ref{tab:gbad-risk-categories} applies the three categories to the same GBAD scenario. It distinguishes the civilian interpretation of harm from the military interpretation and compares a deployed system with the counterfactual case in which the system is not deployed. For category three, this comparison makes explicit that deployment modifies, rather than causes, the external-threat risk.

Risk of the first category may occur due to, for example, the defending missile missing the airborne target and  impacting the ground.
Similarly, harm due to an uncommanded launch of the missile would be counted toward the first category of risk.
With these examples, the dependency on context of the risk evaluation becomes obvious.
An uncommanded launch, for example due to an inadvertent ignition of the missile engine, might result in an impact of the missile into civilian objects, endangering people and goods.
We suspect that this consequence would barely be tolerated by a society during peacetime.
However, assuming a perceived situation of an external threat, the risk balance might shift dramatically.

The challenge with this constellation is obvious when considering the time frame for the design and development of a defense system.
Development cycles in the order of decades are not uncommon.
As a result, the consequences of design decisions are expected to be present in a future that may be shaped by a very different security situation, i.e., an operating context that starkly differs from the assumptions made initially.
An explicit representation of the categories of risk supports the acceptability of systems that need to be effective when the political and security situations demand so.

\section{Conclusion}
\label{sec:conclusion}

Analyzing pre-existing work from both the automotive and defense fields, we find that civilian approaches to safety are only partially applicable to defense.
In order to make defense systems accessible to modern risk management techniques from the civilian world, we formulate the defense system risk balance based on three categories of risk that account for the special needs for achieving security.
Then, proactive risk management can contribute to societal acceptance of defense systems novel to their operating environment which, in turn, is in the interest of security.

Future work remains when it comes to defining a unified taxonomy of security (and, by the way, resilience) that is consistent with existing concepts of safety and risk.
Furthermore, studies on risk perception under various circumstances (wartime versus peacetime) as well as risk acceptance criteria for defense systems in general would be valuable.
The legal aspects should be addressed by future interdisciplinary research as well.
Finally, an integration of the multicategorical risk balance with systems engineering and value-based design seems to be a promising research direction.

\section*{Acknowledgment}
\footnotesize{
We thank the reviewers for their constructive feedback which helped to improve the clarity and focus of this paper.
}

\renewcommand*{\bibfont}{\footnotesize}
\IEEEtriggeratref{6}
\printbibliography 

\end{document}
\typeout{get arXiv to do 4 passes: Label(s) may have changed. Rerun}